\documentclass[lettersize,journal]{IEEEtran}
\usepackage{lipsum}
\usepackage{cuted}
\usepackage{amsfonts}
\usepackage{amssymb}
\usepackage[cmex10]{amsmath}
\usepackage{cite}
\usepackage{graphicx}
\usepackage{verbatim}
\usepackage{subfigure}
\usepackage{bm}
\usepackage{tabularx,booktabs}
\usepackage{algpseudocode}
\usepackage{siunitx}
\usepackage{bm}
\usepackage[ruled]{algorithm2e}
\usepackage{fancyhdr}
\usepackage{setspace}
\usepackage{amsthm}
 \usepackage{enumerate}
 \usepackage{color}
 \usepackage{soul}
\usepackage{multirow}
\usepackage{graphicx}
\usepackage{epstopdf}
\usepackage{threeparttable}
\usepackage{algpseudocode}
\usepackage{amsmath}
\usepackage{wrapfig}
\usepackage{graphicx}

\begin{document}
\renewcommand{\algorithmicrequire}{ \textbf{Input:}}
\renewcommand{\algorithmicensure}{ \textbf{Output:}}
\newenvironment{varalgorithm}[1]
{\algorithm\renewcommand{\thealgorithm}{#1}}
{\endalgorithm}
\newcommand{\be}{\begin{equation}}
\newcommand{\ee}{\end{equation}}
\newcommand{\br}{{\mbox{\boldmath{$r$}}}}
\newcommand{\bp}{{\mbox{\boldmath{$p$}}}}
\newcommand{\bpi}{\mbox{\boldmath{ $\pi $}}}
\newcommand{\bn}{{\mbox{\boldmath{$n$}}}}
\newcommand{\balfa}{{\mbox{\boldmath{$\alpha$}}}}
\newcommand{\ba}{\mbox{\boldmath{$a $}}}
\newcommand{\bta}{\mbox{\boldmath{$\beta $}}}
\newcommand{\bg}{\mbox{\boldmath{$g $}}}
\newcommand{\bPsi}{\mbox{\boldmath{$\Psi $}}}
\newcommand{\bsigma}{\mbox{\boldmath{ $\Sigma $}}}
\newcommand{\bGamma}{{\bf \Gamma }}
\newcommand{\bA}{{\bf A }}
\newcommand{\bP}{{\bf P }}
\newcommand{\bX}{{\bf X }}
\newcommand{\bI}{{\bf I }}
\newcommand{\bR}{{\bf R }}
\newcommand{\bZ}{{\bf Z }}
\newcommand{\bz}{{\bf z }}
\newcommand{\bx}{{\mathbf{x}}}
\newcommand{\bM}{{\bf M}}
\newcommand{\bU}{{\bf U}}
\newcommand{\bD}{{\bf D}}
\newcommand{\bJ}{{\bf J}}
\newcommand{\bH}{{\bf H}}
\newcommand{\bK}{{\bf K}}
\newcommand{\bN}{{\bf N}}
\newcommand{\bC}{{\bf C}}
\newcommand{\bL}{{\bf L}}
\newcommand{\bF}{{\bf F}}
\newcommand{\bv}{{\bf v}}
\newcommand{\bSigma}{{\bf \Sigma}}
\newcommand{\bS}{{\bf S}}
\newcommand{\bs}{{\bf s}}
\newcommand{\bO}{{\bf O}}
\newcommand{\bQ}{{\bf Q}}
\newcommand{\btr}{{\mbox{\boldmath{$tr$}}}}
\newcommand{\bNSCM}{{\bf NSCM}}
\newcommand{\barg}{{\bf arg}}
\newcommand{\bmax}{{\bf max}}
\newcommand{\test}{\mbox{$
\begin{array}{c}
\stackrel{ \stackrel{\textstyle H_1}{\textstyle >} } { \stackrel{\textstyle <}{\textstyle H_0} }
\end{array}
$}}
\newcommand{\tabincell}[2]{\begin{tabular}{@{}#1@{}}#2\end{tabular}}
\newtheorem{Def}{Definition}
\newtheorem{Pro}{Proposition}
\newtheorem{Lem}{Lemma}
\newtheorem{Exa}{Example}
\newtheorem{Rem}{Remark}
\newtheorem{Cor}{Corollary}
\renewcommand{\labelitemi}{$\bullet$}

\title{Automotive Radar Multi-Frame Track-Before-Detect Algorithm Considering Self-Positioning Errors}

\author{Wujun~Li,
Qing Miao,
Ye~Yuan, 
Yunlian~Tian,
Wei~Yi, \emph{Senior Member, IEEE},\\ and
Kah Chan~Teh, \emph{Senior Member, IEEE},
\thanks{This work has been submitted to the IEEE for possible publication. Copyright may be transferred without notice, after which this version may no longer be accessible. }
}



\maketitle

\begin{abstract}
This paper presents a method for the joint detection and tracking of weak targets in automotive radars using the multi-frame track-before-detect (MF-TBD) procedure. Generally, target tracking in automotive radars is challenging due to radar field of view (FOV) misalignment, nonlinear coordinate conversion, and self-positioning errors of the ego-vehicle, which are caused by platform motion. These issues significantly hinder the implementation of MF-TBD in automotive radars. To address these challenges, a new MF-TBD detection architecture is first proposed. It can adaptively adjust the detection threshold value based on the existence of moving targets within the radar FOV. Since the implementation of MF-TBD necessitates the inclusion of position, velocity, and yaw angle information of the ego-vehicle, each with varying degrees of measurement error, we further propose a multi-frame energy integration strategy for moving-platform radar and accurately derive the target energy integration path functions. The
self-positioning errors of the ego-vehicle, which are usually not considered in some previous target tracking approaches, are well addressed.
Numerical simulations and experimental results with real radar data demonstrate large detection and tracking gains over standard automotive radar processing in weak target environments.

\end{abstract}


\section{Introduction}

Autonomous vehicles are crucial components in constructing intelligent transportation systems. To 
realize successful autonomous driving, four basic modules are required: detection, perception, path planning, and control~\cite{Kim_TITS}.
Among these, a reliable and robust perception system for the surroundings of the ego-vehicle is of utmost importance, serving as both a prerequisite and a guarantee for safe driving~\cite {CAO_2024,Cao_TITS}. Automotive radar systems are responsible for the detection of targets and estimating their state including position, velocity, and other higher-order terms. Currently, automotive radar plays an increasingly important role in autonomous driving due to its longer detection range, high measurement accuracy for velocity, high robustness to complex weather conditions, and low cost~\cite{Patole_SPM}.

Target detection and tracking with automotive radars pose challenges when the signal-to-noise ratio (SNR) of the target is low in complex environments.
One aspect is that the detection scenario contains a variety of strong electromagnetic reflectors, such as buildings, road signs, and other metal objects,
especially in urban environments. Additionally, the radar field of view (FOV) changes with translation and rotation of the ego-vehicle. These factors may lead to a low SNR for the target echo due to strong clutter and interference~\cite{Wuhao,Cai_TAES,a10032699}. 
{\color{black}
Another aspect is that transportation environments usually include pedestrians, bicycles, long-distance radar-detected objects, and low-speed vehicles. These targets pose a challenging task for automotive radar due to their relatively low movement velocities and weak echo signal~\cite{Darms_TITS,Patole_SPM,Yan_TGRS}.} 

\begin{figure}
	\centering	\subfigure{\centering\includegraphics[width=3.4in]{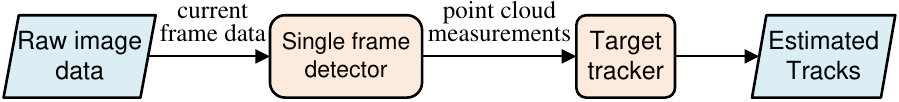}}
	\caption{Traditional DBT scheme in automotive radar systems.}
	\label{fig: DBT}
\end{figure}

Generally, the implementation of target-tracking in automotive radars consists of two procedures: single frame detection (SFD) and tracker, which is also referred to as detect-before-track (DBT)~\cite{orlando2011track,Wuhao,YeHang}, as shown in Fig.~\ref{fig: DBT}. {\color{black}The SFD procedure extracts segmented point groups with sufficient strength at each scan, and these point groups, which exceed the detection threshold, are commonly referred to as point cloud measurements~\cite{Cao_TITS}.}
For surviving detections, target tracking procedures declare the final estimates using the cluster, data association, and filters~\cite{Bar2002Frontmatter,Cao_TITS,Kim_TITS,Kellner_TITS}. However, these approaches have some limitations for weak targets with a low SNR. Specifically, to maintain the tractability of follow-up data association and filters, a higher threshold value is usually set in the SFD stage to eliminate a large number of false noise measurements. This means that target information is also irreversibly discarded after the threshold detection~\cite{Davey2013,wujunTGSR,Hua2023}. Thus, DBT methods are not suitable for weak target detection scenarios in automotive radars.
\begin{figure}
	\centering	\subfigure{\centering\includegraphics[width=3.4in]{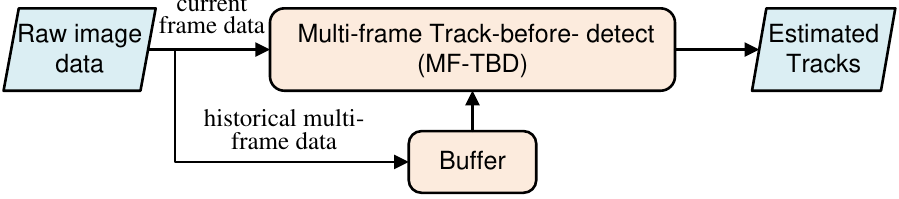}}
	\caption{Potential MF-TBD schemes in automotive radar systems.}
	\label{fig: TBD}
\end{figure}

To improve the detection and tracking performance of weak targets, track-before-detect (TBD) approaches have been proposed~\cite{grossi2013novel,zheng2016Adynamic,Fortin_TITS}. This method directly processes the signal returns of each scan without threshold detection and declares the estimated parameters of targets from raw echo measurements. It can be broadly categorized into single-frame recursive TBD ({\color{black}SFR-TBD}) or multi-frame TBD (MF-TBD). Some classical {\color{black}SFR-TBD} approaches include the histogram probabilistic multi-hypothesis tracker~\cite{Davey2008}, particle filters ~\cite{garcia2012two}, and random finite set algorithms~\cite{Hoseinnezhad2013rfs}, which sequentially estimate the target state over time, similar to the Kalman filter.

MF-TBD can efficiently improve detection and tracking performance for weak targets by jointly processing raw echo data of multiple consecutive frames~\cite{grossi2014track, Zhou2019TAES,Elhoshy2019}, as shown in Fig.~\ref{fig: TBD}. This is largely due to the fact that MF-TBD makes full use of the distinction between targets and noise/clutter in high-dimensional space, and integrates target energy while suppressing the interfering signals. Subsequently, the final estimated target tracks are determined by thresholding the integrated merit function. Some classical implementations of MF-TBD
include Hough transform~\cite{Moyer_Hough}, velocity filtering~\cite{Kennedy_VF,Zhou2019PseudoVFTBD},  and dynamic programming algorithms~\cite{Arnold1993Efficient,Elhoshy2019,Zheng2016}.
In view of the advantages of weak target detection, MF-TBD has been successfully applied to infrared/optical~\cite{Arnold1993Efficient,2002Johnston}, sonar arrays~\cite{Orlando2012}, early warning radar systems~\cite{buzzi2005track,Grossi2016}, and cell tracking in biomedical~\cite{magnusson2015global}. However, the research on MF-TBD in automotive radars is still relatively scarce.

A significant limitation of MF-TBD in automotive radar systems is the substantial degradation of detection and tracking performance caused by the model mismatch from platform motion. {\color{black}A graph based TBD (G-TBD) algorithm was proposed in~\cite{Chen_sensors} to extract target track from multi-frame candidate plots.  This paper assumes that both interest targets and the ego vehicle satisfy the same constant acceleration (CA) model of motion, the yaw angles of the ego vehicle is known and constant during multi-frame measurements. However, the above assumptions do not always hold in practical scenarios of automotive radar.
An MF-TBD procedure for the early detection of moving targets from airborne radars was proposed in~\cite{buzzi2005track} to achieve multi-frame integration using a rough boundary constraint of the velocity or acceleration upper bound. The above approaches eliminate the need to compensate for the motion of the ego vehicle but introduce numerous erroneous tracks during multi-frame energy integration, resulting in performance degradation.} Some straightforward implementations of MF-TBD for airborne radars considering the motion of the platforms were presented at our previous conference article~\cite{Ku_radarcof}. However, these methods overlook nonlinear conversion errors, and 
self-positioning errors in airborne radars, which may lead to the performance loss of algorithms in practice. Automotive radars differ from airborne radars in terms of radar system regimes and detection environments. Furthermore, due to the narrower coverage range of automotive radar beams, issues related to measurement plane misalignment on moving platforms are more pronounced.

\begin{figure}
	\centering	\subfigure{\centering\includegraphics[width=2.2in]{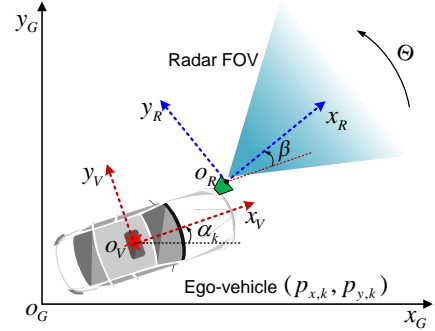}}
	\caption{Automotive radar FOV built in three coordinate systems.}
	\label{fig: Radar}
\end{figure}

In summary, automotive radar MF-TBD also suffers from the following three challenges: \emph{1) Radar FOVs Misalignment}:
automotive radar FOVs vary with the motion of the ego-vehicle.  Most automotive radars use array antennas, these radars suffer from a limited FOV, as shown in Fig.~\ref{fig: Radar}. This gives a potential risk of missing moving targets due to sudden maneuvers or lane changing of the ego-vehicle~\cite{Askeland_TITS}. 
\emph{2) Nonlinear Conversion Errors Among Multiple Coordinate Systems}: automotive radar records the range, azimuth, and velocity information of motion targets relative to the current radar position. All target measurements must be mapped to a uniform coordinate system to track targets. However, this operation includes multiple coordinate conversions and is usually a nonlinear process. 
\emph{3) {\color{black} Self-Positioning Errors (SPE) of the Ego-Vehicle}}: automotive radar target tracking requires position and yaw angle information from the ego-vehicle. This information measured by the navigation system mounted on the ego-vehicle is similarly subject to measurement errors~\cite{Sun_PHD}. However, these errors are usually ignored in the previous target tracking algorithms~\cite{Ku_radarcof,Cao_TITS,Kim_TITS}. 

{\color{black}This paper, building on the conference~\cite{Miao_radarconf}, aims to present a power-efficient MF-TBD algorithm for automotive radars with new detection architecture design and integration path function derivation. Additionally, a detailed performance assessment of algorithms is provided under diverse simulations and experiments using real automotive radar data.} The main contributions of this article are summarized as follows:
\begin{itemize}
\item[1)]\emph{Adaptive Multi-Frame Detection Architecture:}   
We present a novel MF-TBD detection architecture in automotive radars to solve the problem of radar FOV misalignment. It can adaptively adjust the detection threshold value according to the multi-frame integrated path without introducing additional multiple-hypothesis tests. The proposed approach can avoid serious performance deterioration 
when the target suddenly appears or disappears from radar FOV. Moreover, this detection architecture is also suitable for radar FOV misalignment problems of other moving-platform radar systems.
\item[2)]\emph{Accurate Energy Integration of Multi-Frame Measurements:}  
An automotive radar MF-TBD algorithm is proposed to efficiently address radar FOV misalignment, nonlinear conversion errors, and self-positioning errors in automotive radars. Specifically, we propose an automotive radars based MF-TBD algorithm for the case where navigation systems can provide higher precision positioning information for the ego-vehicle, but there are also some measurement errors. In this case, an accurate evolution relationship of the target state is derived, which can provide unbiased and high-confidence search paths by calibrating error terms during multi-frame integration. The self-positioning errors of the ego-vehicle, which are usually not considered in some previous target tracking approaches, are well addressed by accurately deriving the error conversion. The proposed algorithm is further extended to multi-target scenarios. Moreover, the computational complexity of the algorithm is analyzed in practical applications.



\end{itemize}


This paper is organized as follows.  The background and problem statement in terms of 
automotive radars' target detection are given in Section~\ref{se:Background}.
Section~\ref{se:MT-MFD} presents the proposed adaptive MF-TBD detection architecture and   
the implementation of automotive radar
MF-TBD. Simulation results and discussion are given in Section~\ref{sec:Results}, and the experimental results with real data are presented in Section~\ref{sec:Scenario_Experimental}.
Section~\ref{sec:Conclusion} concludes this article.


\section{Background and Problem Statement}
\label{se:Background}
Let us consider a frequency modulated continuous wave (FMCW) radar system mounting on the ego-vehicle horizontally to monitor {\color{black} weak targets on the road, such as pedestrians, bicycles, and low-speed vehicles. A simple schematic is shown in Fig.~\ref{fig: Radar}}. The mounting angle between the {\color{black}$x_\text{R}$}-axis and the {\color{black}$x_\text{V}$}-axis is $\beta$, where the {\color{black}$x_\text{V}$}-axis aligns with the vehicle's longitudinal axis and points forward, and the {\color{black}$x_\text{R}$}-axis is defined along the radar's foresight. 
The radar system's azimuth angle range is $\Theta$. Usually, an ego-vehicle is equipped with $5$ array 
radar systems including $2$ front corner radars, $2$ back corner radars, and $1$ front long-range radar, to cover a $360^{\circ}$ surveillance area~\cite{Patole_SPM,Askeland_TITS}. These radars are mounted on the ego-vehicle with different mounting angles.
This paper mainly considers detecting and tracking arbitrary array radar mounted in the ego-vehicle. The extension of the proposed method to scanning radars 
covering $360^{\circ}$ surveillance area is also feasible, and it does not affect the generality of the subsequent derivations.

Unlike stationary radars, automotive radars always translate and rotate with the ego-vehicle, which means the radar FOVs change over time, as shown in Fig.~\ref{fig: Radar}. 
In order to characterize the kinematics of moving ego-vehicle and moving targets, three different coordinate systems are used~\cite{Kellner_TITS,Cao_TITS}:
\begin{itemize}

\item[1)] \emph{Geodetic Cartesian Coordinate System, {\color{black}$o_\text{G}-x_\text{G}y_\text{G}$}}: The parameters of targets and ego-vehicle (e.g., position, velocity) are built in this coordinate system. The original {\color{black}$o_\text{G}$} is located at a ground observatory;

\item[2)] \emph{Vehicle-body Coordinate System, {\color{black}$o_\text{V}-x_\text{V}y_\text{V}$}}: It is a relative Cartesian coordinate system with the reference point being the ego-vehicle.
The original {\color{black}$o_\text{V}$} is at the centroid of the ego-vehicle;
\item[3)]  \emph{Radar Coordinate System, {\color{black}$o_\text{R}-x_\text{R}y_\text{R}$}}: Radars receive the range and azimuth measurement in the polar coordinate system. The original {\color{black}$o_\text{R}$} is located
at a radar mount point on the ego-vehicle with a mounting angle $\beta$ between the {\color{black}$x_\text{R}$}-axis and the {\color{black}$x_\text{V}$}-axis.
\end{itemize}

\subsection{Motion Model}
\label{se:Motion_Model}
In this section, two motion models need to be considered, including ego-vehicle motion and target motion.

\emph{1) Ego-vehicle Motion Model:} The ego-vehicle is moving in the {\color{black}$o_\text{G}-x_\text{G}y_\text{G}$} coordinate system, and its centroid state vector at time $k$ is ${\color{black}\mathbf{p}}_k=[p_{x,k},p_{\dot{x},_k},p_{y,k},p_{\dot{y},k},\alpha_k]^{\top}$, where $(p_{x,k},p_{y,k})$, $(p_{\dot{x},_k},p_{\dot{y},_k})$ and $\alpha_k$ are centroid position, velocity, and yaw angle of the ego-vehicle, respectively.

Automotive vehicles can use a combination of sensors including a global positioning system (GPS) and inertial measurement unit (IMU) to measure its centroid state ${\color{black}\mathbf{p}}_k$~\cite{Sun_PHD,Kim_TITS}. However, these measurements have some differences from the real values based on factors such as sensor quality, environmental conditions, and sensor fusion algorithm. Given a real centroid state of the ego-vehicle $\bar{{\color{black}\mathbf{p}}}_k$, its corresponding measurement is 
\begin{equation}
	\label{eq:Vehicle_STATE}
	\begin{aligned}
{\color{black}\mathbf{p}}_k=\left[\begin{array}{c}
p_{x,k}\\
p_{\dot{x},k}\\
p_{y,k}\\
p_{\dot{y},k}\\
\alpha_k\end{array}\right]=\left[\begin{array}{c}
\bar{p}_{x,k}\\
\bar{p}_{\dot{x},k}\\
\bar{p}_{y,k}\\
\bar{p}_{\dot{y},k}\\
\bar{\alpha}_k\end{array}\right]+\left[\begin{array}{c}
{\omega}_{x,k}^\text{p}\\
{\omega}_{{\dot{x}},k}^\text{p}\\
{\omega}_{y,k}^\text{p}\\
{\omega}_{\dot{y},k}^\text{p}\\
{\omega}_{\alpha,k}\end{array}\right],
	\end{aligned}
\end{equation}
{\color{black}where 
$\bm{\omega}^\text{p}_{k}=[\,{\omega}_{x,k}^\text{p},{\omega}_{{\dot{x}},k}^\text{p},{\omega}_{y,k}^\text{p},{\omega}_{\dot{y},k}^\text{p},{\omega}_{\alpha,k}]^{\top}$ is usually assumed as the zero-mean, white Gaussian noise~\cite{caron:hal-01509438,Magnusson2012ImprovingAP}, namely $\bm{\omega}^\text{p}_{k}\sim\mathcal{N}(\bm{0},\bm{Q}_{k}^\text{p})$}. Note that the centroid state  ${\color{black}\mathbf{p}}_k$ of the ego-vehicle is assumed known before the radar measurements at time $k$, and its state sequence from the time $1$ to the time $K$, is denoted as ${{P}}_{1:K}=[\,{{\color{black}\mathbf{p}}}_1,\dots,{{\color{black}\mathbf{p}}}_K]$. The quantitative analysis of the correlation between components of the error vector $\bm{\omega}^\text{p}_k$ is complicated because the navigation system usually consists of multiple sensors, i.e., GPU, IMU, and vision sensors.  We assume that, without loss of generality, $\bm{\omega}^\text{p}_k$ is an independent random vector, and its covariance matrix $\bm{Q}_{k}^\text{p}$ is a diagonal array.

\emph{2) Target Motion Model:}  The target kinematic characteristics can be described in a $4$-dimensional (4--D) state space ${\mathbb{R}^{4}}$ of the {\color{black}$o_\text{G}-x_\text{G}y_\text{G}$} coordinates.
The state vector at time $k$ is $\overline{{\color{black}\mathbf{x}}}_k=[\bar{x}_k,\bar{\dot{x}}_k,\bar{y}_k,\bar{\dot{y}}_k]^{\top}\in{\mathbb{R}^{4\times1}}$, where $(\bar{x}_k,\bar{y}_k)$ and $(\bar{\dot{x}}_k,\bar{\dot{y}}_k)$ denote centroid position and velocity, respectively.
{\color{black}According to~\cite{LXR2003}, the  system model for targets can be written as}
\begin{eqnarray}
	\label{eq:CV_STATE}
	\overline{{\color{black}\mathbf{x}}}_{k}&=&\bm{F}_k\overline{{\color{black}\mathbf{x}}}_{k-1}+\bm{C}_k\bm{\omega}^\text{t}_{k-1},\\[1mm]
       \bm{F}_k&=&\bm{I}_2\otimes\left[ \begin{array}{cc}
1&T\\[1mm]
0&1
\end{array} \right],\\
\bm{C}_k&=&\bm{I}_2\otimes\left[ \begin{array}{c}
T^2/2\\
T\end{array} \right],
\end{eqnarray}
where $\bm{\omega}^\text{t}_{k-1}\sim\mathcal{N}(\bm{0},\bm{Q}_{k-1}^\text{t})$ is Gaussian process noise, and $T$ is time interval from the times $k-1$ to the time $k$. {\color{black}The symbol $\otimes$ denotes the Kronecker product.
The term $\bm{I}_2$ is a $2\times2$ identity matrix.} The state sequence of the target over $K$ scans is denoted as $\overline{{X}}_{1:K}=[\overline{{\color{black}\mathbf{x}}}_1,\dots,\overline{{\color{black}\mathbf{x}}}_K]$.

Note that MF-TBD is a batch-processing algorithm with multiple consecutive frames~\cite{WangTAES}, the target motion model of (\ref{eq:CV_STATE}) with a constant velocity (CV) is feasible over a short duration of batch-processing. Besides, extensions to other linear motion models, such as cooperative turning (CT) and CA models, can be done by choosing the right $\overline{{\color{black}\mathbf{x}}}_{k}$, $\bm{F}_k$ and $\bm{C}_k$.

\begin{figure}
	\centering
	\subfigure{\centering\includegraphics[width=2.4in]{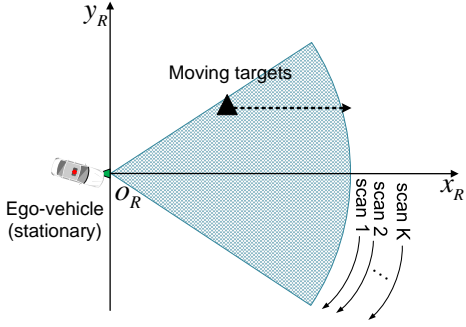}}
	\caption{Radar measurements on a stationary platform.}
	\label{fig: Vehicle_stat}
\end{figure}

\subsection{Radar Measurement Model}
\label{se:Measurement_Model}

Basically, the radar system estimates the relative distance of targets from the measure of the propagation time of a radar beam. Naturally, this sensor acquires range $r$, Doppler velocity $d$, and azimuth $\theta$ information in a polar coordinate about the environment of the ego-vehicle within a distance interval $[r_\text{min},r_\text{max}]$, Doppler velocity range $[d_\text{min},d_\text{max}]$, and azimuth angle range $[\theta_\text{min},\theta_\text{max}]$ with $\Theta=\theta_\text{max}-\theta_\text{min}$. These intervals define the radar FOV.

{\color{black}
Assuming that the raw complex signals received by the FMCW radar
have been processed through the range FFT, Doppler FFT, angle FFT, and the envelope detector~\cite{Patole_SPM}, the resulting output is the range-Doppler-azimuth amplitude data.
The measurement at each time is an $3$-D image data with $N_r\times N_d \times N_{\theta}$ pixels, each with a pixel cell size of  $\Delta_r\times\Delta_d\times\Delta_{\theta}$, where $\Delta_r$, $\Delta_d$, and $\Delta_{\theta}$ represent the discrete resolution cell sizes in range, Doppler velocity, and azimuth dimensions, respectively.
The index set of all pixels
is given by
\begin{equation}
	\label{eq:mea0}
	\begin{aligned}	
\Omega=\big\{(r,d,\theta):\;\;
&r=1,\dots, N_r,\; d=1,\dots, N_d,\; \\
& \theta=1,\dots, N_{\theta}\big\}.
	\end{aligned}
\end{equation}
The amplitude response (the magnitude of I/Q complex signal) in the $(r,d,\theta)$ pixel of the $k$th frame is denoted as $z_k(r,d,\theta)$, and let ${{Z}}_{k}=\{z_k(r,d,\theta),\; (r,d,\theta)\in\Omega\}$ be a stacked set of all pixel responses of the $k$th frame. The target signal at the $(r^*,d^*,\theta^*)$ pixel coming from the target state $\overline{\mathbf{x}}_k$ is denoted by $h_{r^*,d^*,\theta^*}(\overline{\mathbf{x}}_k)$, and $\Omega_k^\text{t}\subset\Omega$ is a stacked set of all pixels coming from the target signal. The target signal could represent the sensor's point spread function (PSF), the target's signature, or a combination of the two~\cite{Davey2012}.
Measurements over $K$ frames are expressed as ${{Z}}_{1:K}=[{{Z}}_{1},\dots,{{Z}}_{K}]$.}

{\color{black}
According to~\cite{Davey2012,Li_tgrs_MT}, the I/Q samples received by radar are commonly modeled as complex Gaussian random variables, its magnitude of the complex signal is Rayleigh distributed under noise-only hypothesis ($H_0$), or Rician distributed under signal-plus-noise
hypothesis ($H_1$). Thus, the probability density function of $z_k(r,d,\theta)$ under $H_0$ is 
\begin{equation}
	\label{eq:H0}
	\begin{aligned}	
    f_{{H}_0}\big(z_{k}(r,d,\theta)\big)&=\frac{z_{k}(r,d,\theta)}{\sigma^2}\exp\left(-\frac{z^2_{k}(r,d,\theta)}{2\sigma^2}\right),
	\end{aligned}
\end{equation}
where $\sigma^2$ denotes the variance of the background noise. 
Since the pixels are assumed to be conditionally independent~\cite{WangTAES}, the probability density function of $z_k(r,d,\theta)$ under $H_1$ is 
\begin{equation}
	\label{eq:H1}
	\begin{aligned}	
    f_{{H}_1}\big(&z_{k}(r,d,\theta);{{\color{black}\mathbf{x}}}_{k},\Omega_k^\text{t}\big)=\prod_{(r,d,\theta)\in\Omega_k^\text{t}}\bigg[\frac{z_{k}(r,d,\theta)}{\sigma^2}\\
&\exp\left(-\frac{z^2_{k}(r,d,\theta)}{2\sigma^2}\right)\mathbf{I}_0\left(\frac{A_{k}(r,d,\theta)z_{k}(r,d,\theta)}{\sigma^2}\right)\bigg],
	\end{aligned}
\end{equation}
where $A_{k}(r,d,\theta), (r,d,\theta)\in\Omega_k^\text{t}$ is the target amplitude of PSF, its center is over the target state and decreases outward, and is assumed to be known. $\mathbf{I}_0(\cdot)$ is the modified Bessel function with order zero. Then, the target SNR quantifies the ratio of peak signal power of the target PSF relative to the noise floor, namely $\text{SNR}=10\log(A^2_{k}/\sigma^2)$ with $A_k=\max\{A_{k}(r,d,\theta), (r,d,\theta)\in\Omega_k^\text{t}\}$, it represents a measure of how easy it is to detect the target~\cite{Davey2012}.
}

\begin{figure}
	\centering
	\subfigure{\centering\includegraphics[width=3.2in]{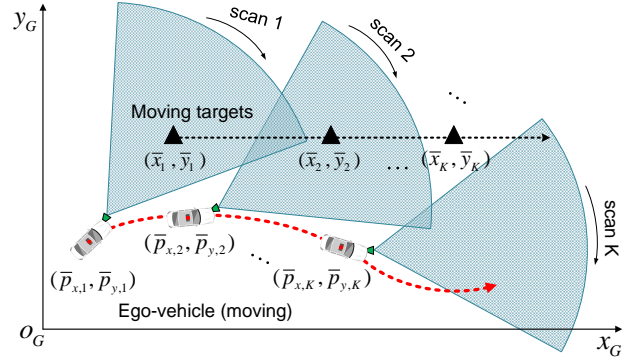}}
	\caption{Radar measurements on a moving platform.}
	\label{fig: Vehicle_mov}
\end{figure}

\subsection{Problem Formulation}
\label{se:Problem}

The aim of MF-TBD is to detect the target from multi-frame echo measurements ${{Z}}_{1:K}$ and declare its estimated state sequence $\widehat{{X}}_{1:K}$ at the same time if the decision is favorable. Following the generalized likelihood ratio test ({\color{black}GLRT}) strategy~\cite{buzzi2008track,orlando2011track}, the target detection through  MF-TBD can be formulated as 
\begin{equation}
	\label{eq:KM_detect1}
\max\limits_{{{X}}_{1:K}\atop{\boldsymbol{x}}_{k}\in\tau({{\color{black}\mathbf{x}}}_{k-1})}\ln{\Lambda}\big({Z}_{1:K};{{X}}_{1:K},\Omega_{1:K}^\text{t}\big)\mathop{\mathrel{\ooalign{\raisebox{.8ex}{$>$}\cr\raisebox{-.8ex}{$<$}}}}\limits_{\text{H}_0}^{H_1}\lambda,
\end{equation}
where 
\begin{itemize}
\item ${\Lambda}(\cdot)$ is the parametric likelihood ratio function, as 
\begin{equation}
	\label{eq:DS1}
	\begin{aligned}
	{\Lambda}\big({Z}_{1:K};{{X}}_{1:K},\Omega_{1:K}^\text{t}\big)
	=\frac{f_{{H}_1}\big({Z}_{1:K};{{X}}_{1:K},\Omega_{1:K}^\text{t}\big)}{f_{{H}_0}\big ({Z}_{1:K}\big)}\\
=\prod_{k=1}^K\prod_{(r,d,\theta)\in\Omega_k^\text{t}}\frac{f_{{H}_1}\big(z_{k}(r,d,\theta);{{\color{black}\mathbf{x}}}_{k},\Omega_k^\text{t}\big)}{f_{{H}_0}\big(z_{k}(r,d,\theta)\big)},
\end{aligned}
\end{equation}
where $f_{{H}_0}(\cdot)$ and $f_{{H}_1}(\cdot)$ are obtained by (\ref{eq:H0}) and  (\ref{eq:H1}), respectively. 
\item $\tau(\cdot)$ is a state transition set defining which destination states at time $k$ may associate with the source state ${{\color{black}\mathbf{x}}}_{k-1}$. Note that, in practice, the maximization of (\ref{eq:KM_detect1}) is carried out in a discrete grid state space of the {\color{black}$o_\text{G}-x_\text{G}y_\text{G}$}  coordinates, i.e., $\mathbb{G}^{4\times1}\in{\mathbb{R}^{4\times1}}$, with a proper quantized criterion $\Delta_x\times\Delta_{\dot{x}}\times\Delta_{y}\times\Delta_{\dot{y}}$, where $\Delta_x$ and $\Delta_y$ are position coordinates, $\Delta_{\dot{x}}$ and $\Delta_{\dot{y}}$ are velocity coordinates in $x$ and $y$ directions, respectively. Then it is needed to enumerate all grid states ${{\color{black}\mathbf{x}}}_{k}\in{\mathbb{G}^{4\times1}}$ to maximize the multi-frame likelihood ratio function.
\item $\lambda$ is a detection threshold to maintain a probability of false alarm $P_{fa}$ that denotes the probability of declaring $H_1$ under the condition that $H_0$ is true~\cite{Cai_TAES}. 
\end{itemize}

We assume that the ego-vehicle is stationary. Echo measurements ${Z}_{1:K}$ and discrete state sequences ${{X}}_{1:K}$ can be modeled in the same  {\color{black}$o_\text{R}-x_\text{R}y_\text{R}$} coordinates, as shown in Fig.~\ref{fig: Vehicle_stat}. In this case, the radar's FOVs during multiple scans 
are completely overlapping and always cover a fixed area of space. Then, some existing MF-TBD methods for ground-based radars~\cite{grossi2013novel,wujunTGSR} can be used to implement (\ref{eq:KM_detect1}). 

However, the moving ego-vehicle makes the implementation of MF-TBD more challenging.
It can be seen from Fig.~\ref{fig: Vehicle_mov} that the radar's FOVs vary with time $k$, and multiple observer coordinate systems are needed to be considered, which increases the difficulty of calculating $\tau(\cdot)$ and $f_{{H}_1}(\cdot)$ in (\ref{eq:KM_detect1}).
The conventional  MF-TBD problem in the stationary-platform radars may no longer represent the actual radars mounted on the motion ego-vehicle. 
The main problems can be summarized as follows:
\begin{itemize}
\item[1)] \emph{Target Loss From Limited Radar FOV:} Clearly, it takes into account not only the movement of the target, but also the movement of the self-vehicle. The target may fall into the outside of FOV in the part times from the 1st scan to the $K$th scan due to the sudden change of vehicles, such as the $K$th scan in Fig.~\ref{fig: Vehicle_mov}. Traditional MF-TBD approaches inevitably suffer performance loss due to missing target measurements during the multi-frame integration, such as $\tau(\cdot)$ is out of radar FOV.

\item[2)] \emph{Nonlinear Mapping Relationship:} As we know, radar measurements are some discrete values with the resolution cell size $\Delta_r\times\Delta_d\times\Delta_{\theta}$, and the volume of the discrete cell gradually increases with the increase of the radial range. These discrete radar measurement plans between different scans are not overlapping owing to the motion of the ego-vehicle. Besides, to obtain the analytic expression of (\ref{eq:DS1}), it also needs to discretize the state vector $\overline{{\color{black}\mathbf{x}}}_{k}$, while the grid state ${{\color{black}\mathbf{x}}}_{k}$ in the coordinate {\color{black}$o_\text{G}-x_\text{G}y_\text{G}$} are not one-to-one correspondence with the radar measurement $z_k(r,d,\theta)$ in the polar coordinates {\color{black}$o_\text{R}-x_\text{R}y_\text{R}$}. The calculation of $f_{{H}_1}(\cdot)$ is difficult.
\item[3)] \emph{Positioning Errors of the Ego-vehicle:} Radar measurements can be mapped to the geodetic Cartesian coordinate system {\color{black}$o_\text{G}-x_\text{G}y_\text{G}$} by compensating the position ambiguity caused by the motion of the ego-vehicle. However, as indicated in Section~\ref{se:Motion_Model}, the obtained centroid state ${\color{black}\mathbf{p}}_k$ of the ego-vehicle unavoidably suffers from measurement errors, and these errors are hard to ignore during the process of MF-TBD. 
\end{itemize}


\section{Automotive Radar MF-TBD Algorithms}
\label{se:MT-MFD}

The analysis of Section~\ref{se:Problem} shows that the automotive radar MF-TBD faces the problems of target loss from radar FOV, nonlinear mapping relationship, and positioning errors of the ego-vehicle. A straightforward application of traditional stationary platform based MF-TBD to automotive radars inevitably results in degraded detection and tracking performance.

In what follows, we first present an adaptive MF-TBD detection architecture to match the sudden target loss from radar FOV during batch processing. 
For the implementation of automotive radar MF-TBD, we further propose a 
multi-frame energy integration strategy,
and accurately derive the multi-frame integration path function with the self-positional errors simultaneously. To solve the multi-target problem,
a direct extension of the proposed algorithm to a multi-target scenario is given by introducing a successive-track-cancellation strategy. Finally, we analyze the computational complexity of the proposed automotive radar MF-TBD algorithm in practical application.


\subsection{Adaptive Multi-Frame Detection Architecture }
\label{se:detection}
Let $\mathcal{M}$ be the set of all possible state sequences that index the paths satisfying the constraints of target kinematics characteristic, and define
\begin{equation}
	\label{eq:possible_track}
	\begin{aligned}	
\mathcal{M}=\big\{{X}_{1:K}:\;\;
&{{\color{black}\mathbf{x}}}_{k}\in\tau({{\color{black}\mathbf{x}}}_{k-1}),\;\; k=2,\dots,K\big\}.
	\end{aligned}
\end{equation}
Given a possible state sequence ${{X}}_{1:K}\in\mathcal{M}$ and multi-frame echo measurements ${{Z}}_{1:K}$,
the state ${{\color{black}\mathbf{x}}}_{k}\in{{X}}_{1:K}$ and the measurement ${{Z}}_{k}$ within radar FOV yield the two possible events at time $k$, as
\begin{equation}
\label{eq:Cartesian_STATA}
	\ell({{\color{black}\mathbf{x}}}_{k},{{Z}}_{k})=\left\{\begin{aligned} &1,\hspace{4mm}\text{target falls into radar FOV},\\
 &0,\hspace{4mm}\text{target does not fall into radar FOV}.\\
 \end{aligned}\right.
\end{equation}
The above formula indicates that moving targets may not always be illuminated by the radar FOV during a batch processing of $K$-frame measurements. When $\ell({{\color{black}\mathbf{x}}}_{k},{{Z}}_{k})=0$, $k=1,\dots,K$,  the detection architecture in  (\ref{eq:KM_detect1}) inevitably suffers from performance loss due to ignoring the fact that target does not fall into radar FOV.



To overcome this drawback, an architecture capable of detecting targets within non-aligned radar FOVs by adaptively designing the threshold value according to $\ell({{\color{black}\mathbf{x}}}_{k},{{Z}}_{k})$ can be developed. Given the state sequence ${{X}}_{1:K}\in\mathcal{M}$, the decision rule (\ref{eq:KM_detect1}) becomes 
\begin{equation}
	\label{eq:KM_detect2}
\max\limits_{{{X}}_{1:K}\in\mathcal{M}}\ln{\Lambda}\big({Z}_{1:K};{{X}}_{1:K},\Omega_{1:K}^\text{t}\big)\mathop{\mathrel{\ooalign{\raisebox{.8ex}{$>$}\cr\raisebox{-.8ex}{$<$}}}}\limits_{\text{H}_0}^{H_{1,\ell}}\lambda_\ell
\end{equation}
\begin{equation}
\label{eq:KM_detect3}
\text{with}~~\ell=\sum_{k=1}^K\ell({{\color{black}\mathbf{x}}}_{k},{{Z}}_{k}),
\end{equation}
where the subscript $\ell$ in the items $H_{1,\ell}$ and $\lambda_\ell$ denotes the total number of frames in which the target falls within the radar FOV. Note that  $\lambda_\ell$ is the detection threshold of $\ell$-frame integration, which can be adaptively adjusted with $\ell$. Specifically, 
solutions of (\ref{eq:KM_detect2}) are given in the following two subsections.

\subsection{Accurate Energy Integration Considering {\color{black}SPE}}
\label{se:Absolute}
\begin{figure}
	\centering	\subfigure{\centering\includegraphics[width=3.2in]{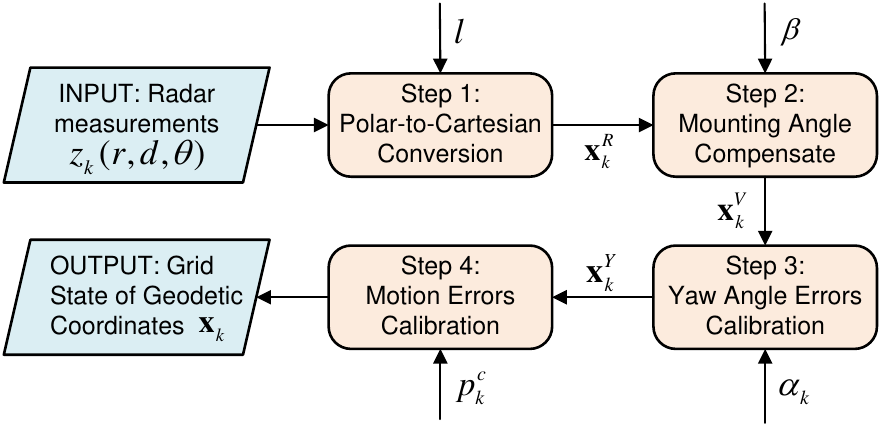}}
	\caption{Operating scheme from the radar measurements to the grid state of geodetic coordinates. }
	\label{fig: Absolute}
\end{figure}

In this subsection, we assume that the higher precision positioning information of the ego-vehicle, such as position, velocity, and yaw angle, can be measured by the navigation system, with some measurement errors. Then, ${X}_{1:K}$ is modeled in an absolute geodetic Cartesian coordinate system {\color{black}$o_\text{G}-x_\text{G}y_\text{G}$}.
Here, the original {\color{black}$o_\text{G}$} can be set as the position of a ground observatory or the position of the ego-vehicle at the initial time with $k=1$.
The basic idea of the proposed automotive radar MF-TBD method is as follows:
The received radar measurements of multiple scans can be first converted to the absolute geodetic Cartesian coordinate system by compensating the platform's motion and its positioning errors. Then,  MF-TBD integrates target energy along the possible state sequence ${X}_{1:K}$ from multi-frame echo measurements. More details are given as follows:

\emph{1) Grid State Modeling Based on {\color{black}SPE}:} To avoid information loss or redundancy in computation due to the matching errors between the Cartesian target state ${\color{black}\mathbf{x}}_{k}$ and the polar measurement $z_{k}(r,d,\theta)$, especially in the case of ego-vehicle movement, we directly define a grid state ${\color{black}\mathbf{x}}_{k}$ through the index information of radar measurement $z_{k}(r,d,\theta)$. This process includes four steps, which are required to complete the coordinate conversion, the mounting angle compensation of radar systems, and yaw angle error and motion error calibration of the ego-vehicle, sequentially, as shown in Fig.~\ref{fig: Absolute}.   

Given $z_{k}(r,d,\theta)$, $k=1,\dots, K$, and a discrete cell of the azimuth velocity $l$, $l=\{1,\dots,\lceil(\dot{\theta}_{\text{max}}-\dot{\theta}_{\text{min}})/\Delta_{\dot{\theta}}\big\rceil\}$ with $\Delta_{\dot{\theta}}=\Delta_{{\theta}}/T$, where $\dot{\theta}_{\text{max}}$ and $\dot{\theta}_{\text{min}}$ denote the maximum and minimum values of the azimuth velocity, respectively, and the operation $\lceil\cdot\rceil$ is the rounding-up function. For convenience, the index set of all 4--D resolution cells is defined as 
\begin{equation}
	\label{eq:mea01}
	\begin{aligned}	
\overline{\Omega}=\big\{(r,d,\theta,l):\;\;
&(r,d,\theta)\in\Omega,\;\; \\
&l=1,\dots,\lceil(\dot{\theta}_{\text{max}}-\dot{\theta}_{\text{min}})/\Delta_{\dot{\theta}}\big\rceil\big\}.
	\end{aligned}
\end{equation}
The time index $k$ is omitted to facilitate the subsequent derivation. 

{\color{black}
Given any grid cell $(r,d,\theta,l)\in\overline{\Omega}$, the measurement state can be calculated by
\begin{equation}
	\label{eq:grid_P0}
	\mathbf{y} =[R,\dot{R},\Theta,\dot{\Theta}]^\top=[r\Delta_r,d_{\text{min}}+d\Delta_d ,\theta\Delta_{{\theta}},\dot{\theta}_{\text{min}}\hspace{-0.2mm}+\hspace{-0.2mm}l\Delta_{\dot{\theta}}]^\top.
\end{equation}
Consider that the radar measurement suffers from system thermal noise and discretization errors. The measurement state $\mathbf{y}$ is usually modeled as a Gaussian random vector~\cite{Fitzgerald,Lerro1993}, which is defined as the linear form of a Gaussian random vector plus a real target state vector, namely
\begin{eqnarray}
	\label{eq:grid_P}
	\mathbf{y} &=&\overline{\mathbf{y}}+\mathbf{n},\;\; \text{with}\;\;\mathbf{n}\sim \mathcal{N}(\bm{0},\bm{M}),\\
    \bm{M}&=&\left[ \begin{array}{cccc}
\sigma_r^2&\rho\sigma_r\sigma_d&0&0\\
\rho\sigma_r\sigma_d&\sigma_d^2&0&0\\
0&0&\sigma_\theta^2&0\\
0&0&0&\sigma_{\dot{\theta}}^2\\
\end{array} \right],
\end{eqnarray}
where $\sigma_r$, $\sigma_d$, $\sigma_\theta$, and $\sigma_{\dot{\theta}}$ are variances of the errors in range, Doppler velocity, azimuth, and azimuth velocity dimensions, respectively.  
The notation $\rho$ represents the correlation coefficient that characterizes the range-Doppler coupling of radar chirp waveforms~\cite{Fitzgerald}, and it is determined by the parameters of the waveform.}

Without loss of generality, a  grid state ${\color{black}\mathbf{x}}^\text{R}$ in radar Cartesian coordinates can be obtained from the measurement state  $\mathbf{y}$, and it can be further decomposed as
linear transformations of a  random vector plus a real target state vector $\overline{{\color{black}\mathbf{x}}}^\text{R}$, namely
\begin{equation}
\label{eq:Cartesian_STATA}
{\color{black}
\begin{aligned}
	\mathbf{x}^\text{R} =s(\mathbf{y})&=\left[ \begin{array}{c}
R \cos\Theta\\
\dot{R}\cos\Theta -R  \dot{\Theta}\sin\Theta \\
R \sin{\Theta}\\
\dot{R} \sin\Theta  +R  \dot{\Theta}\cos\Theta \\
\end{array}\right]\\[1mm]
&= \overline{{\color{black}\mathbf{x}}}^\text{R}+{\color{black}\mathbf{v}}^\text{R},
\end{aligned}}
\end{equation}
where ${\color{black}\mathbf{v}}^\text{R}$ is the converted measurement error from the polar coordinates to Cartesian coordinates. Following the lead of~\cite{Lerro1993,Bordonaro2014}, ${\color{black}\mathbf{v}}^\text{R}$ can be approximated by Gaussian random vector with ${\color{black}\mathbf{v}}^\text{R}\sim\mathcal{N}(\bm{\mu}^\text{R},\bm{Q}^\text{R})$, whose explicit expressions about the mean $\mathbb{E}[{\color{black}\mathbf{v}}^\text{R}]=\bm{\mu}^\text{R}$ and the covariance matrix $\text{Var}({\color{black}\mathbf{v}}^\text{R})=\bm{Q}^\text{R}$ can be found in~\cite{Bordonaro2014}.

For the grid state ${\color{black}\mathbf{x}}^\text{R}$ in radar Cartesian coordinates, it can be converted to  the vehicle-body coordinates by rotating $\beta$ degrees clockwise, as 
\begin{eqnarray}
	\label{eq:CV_STATE_grid}
	{\color{black}\mathbf{x}}^\text{V} &=&\bm{R}(\beta){\color{black}\mathbf{x}}^\text{R}\hspace{2mm}=\hspace{2mm}\bm{R}(\beta)  \overline{{\color{black}\mathbf{x}}}^\text{R}+\bm{R}(\beta) {\color{black}\mathbf{v}}^\text{R},\\[0.2mm] \label{eq:CV_STATE_I}
            \bm{R}(\beta)&=&\left[ \begin{array}{cccc}
\cos\beta&0&\sin\beta&0\\
0&\cos\beta&0&\sin\beta\\
-\sin\beta&0&\cos\beta&0\\
0&-\sin\beta&0&\cos\beta\\
\end{array} \right],
\end{eqnarray}
where $\beta$ is a fixed mounting angle.  Let  ${\color{black}\mathbf{v}}^\text{V}=\bm{R}(\beta) {\color{black}\mathbf{v}}^\text{R}$ denote  the conversion error,  and its mean and covariance matrix satisfy 
\begin{eqnarray}
	\bm{\mu}^\text{V}\hspace{2mm}= &\mathbb{E}\left[{\color{black}\mathbf{v}}^\text{V}\right]&=\hspace{2mm}\bm{R}(\beta)\bm{\mu}^\text{R},	\label{eq:mean_cov1_1}\\[1mm]    \bm{Q}^\text{V}\hspace{2mm}=&\text{Var}\left({\color{black}\mathbf{v}}^\text{V}\right)&=\hspace{2mm}\bm{R}(\beta)\bm{Q}^\text{R}\bm{R}^{\top}(\beta).	\label{eq:mean_cov1_2}
\end{eqnarray}

To obtain the grid state ${\color{black}\mathbf{x}}$ in geodetic Cartesian coordinates, it is needed to compensate for the deviation term ${\color{black}\mathbf{p}}$ due to the motion of the ego-vehicle. Note that the centroid state ${\color{black}\mathbf{p}}$ of the ego-vehicle usually suffers the measurement error $ \bm{\omega}^\text{p}$. Besides, ${\color{black}\mathbf{p}}$ is assumed to be known before radar measurements.
Thus, it is reasonable to assume that the centroid state ${\color{black}\mathbf{p}}$ and the grid state ${\color{black}\mathbf{x}}^\text{V} $ are independent of each other.
Similar to (\ref{eq:CV_STATE_I}), the grid state after calibrating the yaw angle error $\alpha$ can be expressed as 
\begin{equation}
	\label{eq:grid_STATE_Y}
	{\color{black}\mathbf{x}}^\text{Y} =\bm{R}(\alpha){\color{black}\mathbf{x}}^\text{V}.
\end{equation}
Then the above equation can be further decomposed as ${\color{black}\mathbf{x}}^\text{Y}=\overline{{\color{black}\mathbf{x}}}^\text{Y}+{\color{black}\mathbf{v}}^\text{Y}$, where $\overline{{\color{black}\mathbf{x}}}^\text{Y}$ is the real target state, and ${\color{black}\mathbf{v}}^\text{Y}$
is a conversion error.

By substituting $\alpha=\bar{\alpha}+{\omega}_{\alpha}$ and (\ref{eq:CV_STATE_grid})--(\ref{eq:mean_cov1_2}) into (\ref{eq:grid_STATE_Y}), and after some algebraic
manipulations,  the mean of the error ${\color{black}\mathbf{v}}^\text{Y}$ conditioned on $\alpha$ and ${\color{black}\mathbf{x}}^\text{V}=[x^\text{v},\dot{x}^\text{v},y^\text{v},\dot{y}^\text{v}]^{\top}$ is  
\begin{equation}
	\label{eq:grid_STATE_Y1}
\begin{aligned}	\bm{\mu}^\text{Y}&=\mathbb{E}\left[{\color{black}\mathbf{v}}^\text{Y}\right] 
 =\mathbb{E}\left[{\color{black}\mathbf{x}}^\text{Y}-\overline{{\color{black}\mathbf{x}}}^\text{Y}\right]\\[1mm]
 &=\mathbb{E}\left[\bm{R}(\alpha){\color{black}\mathbf{x}}^\text{V}-\bm{R}(\alpha-{\omega}_{\alpha})({{\color{black}\mathbf{x}}}^\text{V}-{\color{black}\mathbf{v}}^\text{V})\right]\\[1mm]
 &=\mathbb{E}\left[\begin{array}{c}
   \begin{aligned}
   &x^\text{v}\cos\alpha
   +y^\text{v}\sin\alpha+\lambda_\alpha\left( \bm{\mu}^\text{V}_1-x^\text{v}\right)\cos\alpha \\
   &+\lambda_\alpha\left( \bm{\mu}^\text{V}_3-y^\text{v}\right)\sin\alpha\end{aligned} \\
      \begin{aligned}&\dot{x}^\text{v}\cos\alpha
   +\dot{y}^\text{v}\sin\alpha+\lambda_\alpha\left( \bm{\mu}^\text{V}_2-\dot{x}^\text{v}\right)\cos\alpha \\
   &+\lambda_\alpha\left( \bm{\mu}^\text{V}_4-\dot{y}^\text{v}\right)\sin\alpha
   \end{aligned} \\
      \begin{aligned}&y^\text{v}\cos\alpha
   -x^\text{v}\sin\alpha+\lambda_\alpha\left(x^\text{v} -\bm{\mu}^\text{V}_1\right)\sin\alpha \\
   &+\lambda_\alpha\left( \bm{\mu}^\text{V}_3-y^\text{v}\right)\cos\alpha\end{aligned} \\
    \begin{aligned}&\dot{y}^\text{v}\cos\alpha
   -\dot{x}^\text{v}\sin\alpha+\lambda_\alpha\left(\dot{x}^\text{v} -\bm{\mu}^\text{V}_2\right)\sin\alpha \\
   &+\lambda_\alpha\left( \bm{\mu}^\text{V}_4-\dot{y}^\text{v}\right)\cos\alpha\end{aligned} \\
 \end{array}\right],
\end{aligned}
\end{equation}
where $\lambda_\alpha=\exp^{-{Q}^\text{p}_{55}/2}$ with ${Q}^\text{p}_{55}$ being the element in row $5$ and column $5$ of the matrix $\bm{Q}^\text{p}$. Note that $\bm{Q}^\text{p}$ has been defined in (\ref{eq:Vehicle_STATE}), but the time index $k$ of the subscript has been omitted for notational convenience.
Similarly, $\bm{\mu}^\text{V}_i$, $i=1,\dots,4$ is the $i$th element in the vector $\bm{\mu}^\text{V}$. Following that,
the covariance of ${\color{black}\mathbf{v}}^\text{Y}$ is given by 
\begin{equation}
\label{eq:carti_cov}
\begin{aligned}
\bm{Q}^\text{Y}=\text{Var}\left({\color{black}\mathbf{v}}^\text{Y}\right)
=
\left[ \begin{array}{cccc}
{Q}_{11}^\text{Y}&{Q}_{12}^\text{Y}&{Q}^\text{Y}_{13}&{Q}^\text{Y}_{14}\\
{Q}_{21}^\text{Y}&{Q}^\text{Y}_{22}&{Q}^\text{Y}_{23}&{Q}^\text{Y}_{24}\\
{Q}^\text{Y}_{31}&{Q}^\text{Y}_{32}&{Q}^\text{Y}_{33}&{Q}^\text{Y}_{34}\\
{Q}^\text{Y}_{41}&{Q}^\text{Y}_{42}&{Q}^\text{Y}_{43}&{Q}^\text{Y}_{44}\\
\end{array}\right],
\end{aligned}
\end{equation}
where the explicit expression of each item in  the matrix $\bm{Q}^\text{Y}$ is given in Appendix~\ref{Apen: cov}.

Next, we further calibrate the motion errors of the ego-vehicle, i.e.,  ${\color{black}\mathbf{p}}^\text{c}=\left[p_{x},p_{\dot{x},},p_{y},p_{\dot{y}}\right]^{\top}\in{\color{black}\mathbf{p}}$,
to obtain the grid state ${\color{black}\mathbf{x}}$,
\begin{equation}
	\label{eq:grid_STATE1}
	{\color{black}\mathbf{x}} ={\color{black}\mathbf{x}}^\text{Y}+{\color{black}\mathbf{p}}^\text{c}= \overline{{\color{black}\mathbf{x}}}^\text{Y}+{\color{black}\mathbf{v}}^\text{Y}+\overline{{\color{black}\mathbf{p}}}^\text{c}+\bm{\omega}^\text{c},
\end{equation}
where $\bm{\omega}^\text{c}=[{\omega}_{x,k}^\text{p},{\omega}_{{\dot{x}}}^\text{p},{\omega}_{y}^\text{p},{\omega}_{\dot{y}}^\text{p}]^{\top}\in\bm{\omega}^{p}$ denotes the position and velocity errors.  Let $\overline{{\color{black}\mathbf{x}}}=\overline{{\color{black}\mathbf{x}}}^\text{Y}+\overline{{\color{black}\mathbf{p}}}^\text{c}$ and ${\color{black}\mathbf{v}}={\color{black}\mathbf{v}}^\text{Y}+\bm{\omega}^\text{c}$ denote the real grid state and its conversion error, respectively. Based on (\ref{eq:Vehicle_STATE}), (\ref{eq:grid_STATE_Y1}) and (\ref{eq:carti_cov}), the mean and  covariance matrix of ${\color{black}\mathbf{v}}$ can be written as 
\begin{eqnarray}
	\bm{\mu}&=& \mathbb{E}[{\color{black}\mathbf{v}}]\hspace{3.2mm}=\hspace{2mm}\bm{\mu}^\text{Y},	\label{eq:grid_mean_1}\\[1mm]
        \bm{Q}&=&\text{Var}({\color{black}\mathbf{v}})=\hspace{2mm}\bm{Q}^\text{Y}+\bm{Q}^\text{c},	\label{eq:grid_cov_2}
\end{eqnarray}
where $\bm{Q}^\text{c}$ denotes the covariance matrix of $\bm{\omega}^\text{c}$, which has been given in (\ref{eq:Vehicle_STATE}).

\emph{2) Search Path Design:} 
With the above derivations in mind, the possible search path of automotive radar MF-TBD can be derived as follows.

Equation (\ref{eq:grid_STATE1}) indicates that, for any grid state ${\color{black}\mathbf{x}}_k\in{\mathbb{G}_k^{4\times1}}$ at time $k$, $k=1,\dots,K$, the following  equation holds
\begin{equation}
	{\color{black}\mathbf{x}}_k= \overline{{\color{black}\mathbf{x}}}_k+{\color{black}\mathbf{v}}_k,	\label{eq:grid_state}
\end{equation}
where the mean ${\color{black}\mathbf{v}}_k$ and covariance matrix $\bm{Q}_k$ of the conversion error  can be calculated using  (\ref{eq:grid_mean_1}) and (\ref{eq:grid_cov_2}), respectively. Note that
the grid state space ${\mathbb{G}_k^{4\times1}}$ changes with the radar FOV since ${\color{black}\mathbf{x}}_k$ is obtained through  $z_{k}(r,d,\theta)$ according to (\ref{eq:Cartesian_STATA}), as shown in Fig.~\ref{fig: Absolute}.

By substituting real target state (\ref{eq:CV_STATE}) into (\ref{eq:grid_state}), we have 
\begin{equation}
\label{eq:grid_state_predict}
\begin{aligned}
	{\color{black}\mathbf{x}}_k\hspace{2mm}&= \hspace{2mm}\bm{F}_k\overline{{\color{black}\mathbf{x}}}_{k-1}+\bm{C}_k\bm{\omega}^\text{t}_{k-1}+{\color{black}\mathbf{v}}_k\\
 &=\hspace{2mm}\bm{F}_k({{\color{black}\mathbf{x}}}_{k-1}-{\color{black}\mathbf{v}}_{k-1})+\bm{C}_k\bm{\omega}^\text{t}_{k-1}+{\color{black}\mathbf{v}}_k\\
  &=\hspace{2mm}\bm{F}_k{{\color{black}\mathbf{x}}}_{k-1}-\bm{F}_k{\color{black}\mathbf{v}}_{k-1}+\bm{C}_k\bm{\omega}^\text{t}_{k-1}+{\color{black}\mathbf{v}}_k.
\end{aligned}
\end{equation}

We define $\bm{\omega}^\text{t}_{k|k-1}=-\bm{F}_k{\color{black}\mathbf{v}}_{k-1}+\bm{C}_k\bm{\omega}^\text{t}_{k-1}+{\color{black}\mathbf{v}}_k$ as the prediction error. By combining (\ref{eq:CV_STATE}), (\ref{eq:grid_mean_1}) and (\ref{eq:grid_cov_2}), the mean and covariance matrix of $\bm{\omega}^\text{t}_{k|k-1}$ satisfy
\begin{eqnarray}
	\bm{\mu}_{k|k-1}&=&\mathbb{E}\big[\bm{\omega}^\text{t}_{k|k-1}\big]\nonumber\\
 &=&-\bm{F}_k\bm{\mu}_{k-1}^\text{Y}+\bm{\mu}_{k}^\text{Y},
 \label{eq:grid_state_predict_M_C1}\\[1mm]
\bm{Q}_{k|k-1}&=&\text{Var}\big(\bm{\omega}^\text{t}_{k|k-1}\big)\nonumber\\ 
&=& \bm{F}_k\left(\bm{Q}^\text{Y}_{k-1}+\bm{Q}^\text{c}_{k-1}\right)\bm{F}_k^{\top}+\bm{C}_k\bm{Q}_{k-1}^\text{t}\bm{C}_k^{\top}\nonumber\\
&&+\bm{Q}^\text{Y}_{k}+\bm{Q}^\text{c}_{k}.
\label{eq:grid_state_predict_M_C2}
\end{eqnarray}

Now, given the grid states ${\color{black}\mathbf{x}}_k$ and ${\color{black}\mathbf{x}}_{k-1}$, the possible search path between adjacent frames can be defined through the Mahalanobis distance, as  
\begin{equation}
 \label{eq:grid_state_possible}
\begin{aligned}
&\tau_\text{{\color{black}SPE}}({\color{black}\mathbf{x}}_{k-1})=\Big\{{\color{black}\mathbf{x}}_k\in{\mathbb{G}_k^{4\times1}}:\left( {\color{black}\mathbf{x}}_k-\bm{F}_k{{\color{black}\mathbf{x}}}_{k-1}-\bm{\mu}_{k|k-1}\right)^\top\\
&\hspace{10mm}\bm{Q}_{k|k-1}^{-1}\left( {\color{black}\mathbf{x}}_k-\bm{F}_k{{\color{black}\mathbf{x}}}_{k-1}-\bm{\mu}_{k|k-1}\right)\leq \varUpsilon_{\text{{\color{black}SPE}}}
\Big\},
\end{aligned}
\end{equation}
where $\varUpsilon_{\text{{\color{black}SPE}}}$ is a predefined threshold value to limit the size of the set of possible searching states.
The subscript $\text{{\color{black}SPE}}$ in $\varUpsilon_{\text{{\color{black}SPE}}}$ and $\tau_\text{{\color{black}SPE}}(\cdot)$ indicates that the enumerated search is built on {\color{black}SPE}. Similarly, we can define an inverse function $\tau^{-1}_\text{{\color{black}SPE}}({\color{black}\mathbf{x}}_{k})$ to ascertain which state ${\color{black}\mathbf{x}}_{k-1}\in{\mathbb{G}_{k-1}^{4\times1}}$ at time $k-1$ may associate with ${\color{black}\mathbf{x}}_{k}$. 

\emph{3) {\color{black}SPE Based MF-TBD (SPE-MF-TBD)}:} 
Finally, we substitute (\ref{eq:grid_state_possible}) into (\ref{eq:KM_detect2}) to construct the test problem of MF-TBD. According to the dynamic programming algorithm~\cite{Davey2008,barniv1985dynamic}, the multi-frame maximization of (\ref{eq:KM_detect2}) can be decomposed into multiple iteratively solved sub-optimization problems.
Define the integrated merit-function of the grid state ${{\color{black}\mathbf{x}}}_{k}$ at time $k$ as 
\begin{eqnarray}
	\label{eq:KM_detect1_AMM}
\mathcal{I}({{\color{black}\mathbf{x}}}_{k})\hspace{-2mm}&=&\hspace{-4mm}\max\limits_{{{X}}_{1:k-1}\atop{{\color{black}\mathbf{x}}}_{n-1}\in\tau_\text{{\color{black}SPE}}^{-1}({{\color{black}\mathbf{x}}}_{n})}\sum_{n=1}^k\sum_{(r,d,\theta)\in\Omega_n^\text{t}}\hspace{-2mm}\log{\Lambda}\big(z_{n}(r,d,\theta);{{\color{black}\mathbf{x}}}_{n},\Omega_n^\text{t}\big)\notag\\
&=&\max\limits_{{{X}}_{1:k-1}\atop{{\color{black}\mathbf{x}}}_{n-1}\in\tau_\text{{\color{black}SPE}}^{-1}({{\color{black}\mathbf{x}}}_{n})}\sum_{n=1}^k\Lambda({{\color{black}\mathbf{x}}}_{n}).
\end{eqnarray}
Note that $\mathcal{I}({{\color{black}\mathbf{x}}}_{1})=\Lambda({{\color{black}\mathbf{x}}}_{1})$  denotes the  merit-function of the initial state ${\color{black}\mathbf{x}}_{1}\in{\mathbb{G}_{1}^{4\times1}}$. At this time, only the states within radar FOV are considered with $\ell({{\color{black}\mathbf{x}}}_{1},{{Z}}_{1})=1$. 

\begin{figure}
	\centering
\subfigure{\centering\includegraphics[width=3.5in]{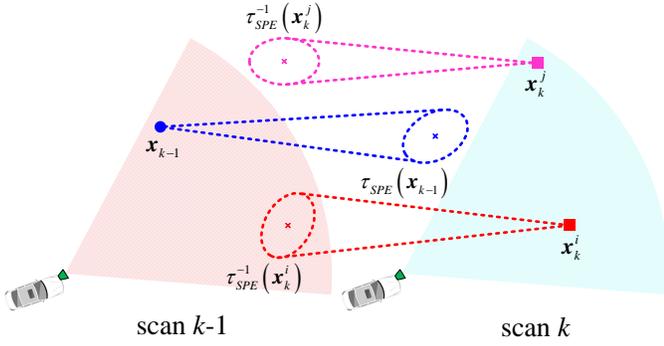}}
	\caption{Update and transmission schematic of {\color{black}SPE-MF-TBD}.}
	\label{fig: FOV_Mismatch}
\end{figure}

To avoid  target loss caused by  non-aligned radar FOV during multi-frame integration, we not only update the merit-function of   destination states ${\color{black}\mathbf{x}}_{k}\in{\mathbb{G}_{k}^{4\times1}}$, but also transmit the merit-function of  source states ${\color{black}\mathbf{x}}_{k-1}\in{\mathbb{T}_{k-1}}\in{\mathbb{G}_{k-1}^{4\times1}}$ in the remaining state set of the non-aligned FOV, as
\begin{equation}
	\label{eq:satte}
{\mathbb{T}_{k-1}}={\mathbb{G}_{k-1}^{4\times1}}~\big\backslash~\bigcup_{{{\color{black}\mathbf{x}}}_{k}\in{\mathbb{G}_{k}^{4\times1}}}\tau^{-1}_\text{{\color{black}SPE}}({\color{black}\mathbf{x}}_{k}),
\end{equation}
where $\cup$ denotes the union operations.

\emph{a) Update Stage:} Now, for $k=2,\dots,K$, the merit-function of any state ${\color{black}\mathbf{x}}_{k}\in{\mathbb{G}_{k}^{4\times1}}$ with $\ell({{\color{black}\mathbf{x}}}_{k},{{Z}}_{k})=1$ can be sequentially updated through the following two cases
\begin{itemize}
\item If $\exists {{\color{black}\mathbf{x}}}_{k-1}\in\tau_\text{{\color{black}SPE}}^{-1}({{\color{black}\mathbf{x}}}_{k})$, ${{\color{black}\mathbf{x}}}_{k-1}\in{\mathbb{G}_{k-1}^{4\times1}}$, we have 
\begin{align}
\mathcal{I}({{\color{black}\mathbf{x}}}_{k})&=\max\limits_{{{\color{black}\mathbf{x}}}_{k-1}\in\tau_\text{{\color{black}SPE}}^{-1}({{\color{black}\mathbf{x}}}_{k})}\mathcal{I}({{\color{black}\mathbf{x}}}_{k-1})+\Lambda({{\color{black}\mathbf{x}}}_{k})\label{eq:KM_detect1_AMM1},\\[1mm]
\psi({{\color{black}\mathbf{x}}}_{k})&=\arg\max\limits_{{{\color{black}\mathbf{x}}}_{k-1}\in\tau_\text{{\color{black}SPE}}^{-1}({{\color{black}\mathbf{x}}}_{k})}\hspace{-2mm}\mathcal{I}({{\color{black}\mathbf{x}}}_{k-1}),\label{eq:KM_detect1_AMM2}
\end{align}
where $\psi({{\color{black}\mathbf{x}}}_{k})$ is a backtracking function of tracks used to conserve the historical states. A  schematic diagram is given in Fig.~\ref{fig: FOV_Mismatch} to illustrate that search status falls within the radar FOV, such as ${{\color{black}\mathbf{x}}}_{k}^i$.
\item If $\forall {{\color{black}\mathbf{x}}}_{k-1}\in\tau_\text{{\color{black}SPE}}^{-1}({{\color{black}\mathbf{x}}}_{k})$, ${{\color{black}\mathbf{x}}}_{k-1}\notin{\mathbb{G}_{k-1}^{4\times1}}$, we have 
\begin{align}
\mathcal{I}({{\color{black}\mathbf{x}}}_{k})&=\Lambda({{\color{black}\mathbf{x}}}_{k})\label{eq:KM_detect1_AMM1},\\[1mm]
\psi({{\color{black}\mathbf{x}}}_{k})&=0.\label{eq:KM_detect1_AMM22}
\end{align}
Note that the new merit-function is initialized during multi-frame integration when the search status falls outside the radar FOV, such as ${{\color{black}\mathbf{x}}}_{k}^j$ in Fig.~\ref{fig: FOV_Mismatch}. 
\end{itemize}

\emph{b) Transmission Stage:} After updating the integrated merit-function of all states in the set ${\mathbb{G}_{k}^{4\times1}}$, the merit-function of the remaining state  ${\color{black}\mathbf{x}}_{k-1}\in{\mathbb{T}_{k-1}}$ at the  time $k-1$ is transmitted to the time $k$, 
\begin{align}
\mathcal{I}({{\color{black}\mathbf{x}}}_{k})&=\mathcal{I}({{\color{black}\mathbf{x}}}_{k-1}) ~\text{with}~\ell({{\color{black}\mathbf{x}}}_{k},{{Z}}_{k})=0, \label{eq:KM_detect1_AMM1_update}
\\[1mm]
\psi({{\color{black}\mathbf{x}}}_{k})&={{\color{black}\mathbf{x}}}_{k-1},~{{\color{black}\mathbf{x}}}_{k}=\bm{F}_k{{\color{black}\mathbf{x}}}_{k-1},
\end{align}
such as ${{\color{black}\mathbf{x}}}_{k-1}$ in Fig.~\ref{fig: FOV_Mismatch}.
Simultaneously, the transmitted state is stored in the set ${\mathbb{G}_{k}^{4\times1}}$, 
\begin{equation}
	\label{eq:sattesss}
{\mathbb{G}_{k}^{4\times1}}= {\mathbb{G}_{k}^{4\times1}}~\cup~\{{{\color{black}\mathbf{x}}}_{k}:~{{\color{black}\mathbf{x}}}_{k}=\bm{F}_k{{\color{black}\mathbf{x}}}_{k-1},{{\color{black}\mathbf{x}}}_{k-1}\in{\mathbb{T}_{k-1}}\}.
\end{equation}
Following that, the merit-function $\mathcal{I}({{\color{black}\mathbf{x}}}_{K})$ and $\ell=\sum_{k=1}^K\ell({{\color{black}\mathbf{x}}}_{k},{{Z}}_{k})$ at the final stage can be obtained. The estimated state sequence $\widehat{{X}}_{1:K}$ is declared by 
\begin{eqnarray}
\widehat{{\color{black}\mathbf{x}}}_{K}&=&\arg\max\limits_{{{\color{black}\mathbf{x}}}_{K}\in{\mathbb{G}_K^{4\times1}}}\mathcal{I}({{\color{black}\mathbf{x}}}_{K}),\; \text{ s.t. }   \mathcal{I}({{\color{black}\mathbf{x}}}_{K})> \lambda_\ell,	\label{eq:KM_detect1_AMM_dec}\\[1mm]
\widehat{{\color{black}\mathbf{x}}}_{k-1}&=&\psi(\widehat{{\color{black}\mathbf{x}}}_{k}), \; \hspace{10mm}k=K,K-1,\dots,2.\label{eq:KM_detect1_AMM_out}
\end{eqnarray}

\begin{Rem}
Target loss caused by the non-aligned radar FOV, as shown in Fig.~\ref{fig: Vehicle_mov}, can be well resolved through the introduction of update and transmission steps during multi-frame integration, and the design of the adaptive detection threshold $\lambda_\ell$.
\end{Rem}
\begin{Rem} To achieve automotive radar MF-TBD, we need to consider multiple coordinate conversions, as shown in Fig.~\ref{fig: Absolute}. Consequently, the conversion error from the nonlinear mapping function will no longer be zero mean, it causes the grid state used to calculate possible search paths to become biased.  By accurately deriving the mean and covariance matrix of the conversion error, the proposed {\color{black}SPE-MF-TBD} algorithm can provide an unbiased and high-confidence state search region after offsetting the deviation term of the errors. Furthermore, the detection and tracking performance of MF-TBD can be improved.
\end{Rem}
\begin{Rem} Some existing target tracking approaches in automotive radars usually assume that the navigation system can obtain the real centroid state of the ego-vehicle, $\bar{{\color{black}\mathbf{p}}}_k$. However, this assumption is not held in practice due to measurement errors. The proposed {\color{black}SPE-MF-TBD} relaxes the assumptions to the more general case.
\end{Rem}

\subsection{Multi-Target Extraction}
\label{se:MT}
The proposed {\color{black}SPE-MF-TBD} algorithm mainly considers a single target detection scenario. Thus, the state sequence $\widehat{{X}}_{1:K}$ with the maximum integration merit-function is the estimated target track according to (\ref{eq:KM_detect1_AMM_dec}) and (\ref{eq:KM_detect1_AMM_out}).
An extension of the proposed algorithm to multi-target scenarios is straightforward via the introduction of a successive-track-cancellation (STC) strategy~\cite{yi2013efficient,buzzi2008track}. Let $\mathbb{D}_\ell$ denote the set of tracks with the $\ell$-frame integration, whose final stage merit-function exceeds the detection threshold $\lambda_\ell$, as 
\begin{equation}
\label{eq:threshold}
\begin{aligned}
&\mathbb{D}_\ell=\big\{{{X}}_{1:K}:~~\mathcal{I}({{\color{black}\mathbf{x}}}_{K})> \lambda_\ell,~~{{\color{black}\mathbf{x}}}_{K}\in{\mathbb{G}_{K}^{4\times1}}
\big\}.
\end{aligned}
\end{equation}
Then, all tracks are collected into the set $\mathbb{D}=\{\mathbb{D}_\ell,\ell=1,\dots,K\}$.

For each $\mathbb{D}_\ell\in\mathbb{D}$ and $\mathbb{C}_\ell=\emptyset$,  the track with the  maximum  merit-function within the set $\mathbb{D}$ is extracted, as 
\begin{equation}
\label{eq:threshold}
\begin{aligned}
&\mathbb{C}_\ell=\mathbb{C}_\ell \cup\widehat{{X}}_{1:K},~~\widehat{{X}}_{1:K}=\arg\max\limits_{{{X}}_{1:K}\in\mathbb{D}_\ell}\mathcal{I}({{\color{black}\mathbf{x}}}_{K}).
\end{aligned}
\end{equation}
Then, all tracks within $\mathbb{D}_\ell$ that share $L$, $L=1,\dots,K$, common measurements with $\widehat{{X}}_{1:K}$ are deleted from the set $\mathbb{D}_\ell$. We repeat the above steps until the set of $\mathbb{D}_\ell$ is empty. Finally, all tracks in sets $\{\mathbb{C}_\ell,\ell=1,\dots,K\}$  are declared.

\subsection{Computational Complexity Analysis}
\label{se:Complexity}
{\color{black}In this subsection, approximate analyses are given to illustrate the computational complexities of the traditional DBT and the proposed {\color{black}SPE-MF-TBD}.

\emph{1) DBT Method:} It is shown in Fig.~\ref{fig: DBT} that DBT iteratively processes the single frame measurement ${{Z}}_{k}$. The computational complexity of DBT mainly consists of the detector and tracker. Assume that the detection threshold of the cell under test is calculated from $\xi$ reference units around it, the complexity of completing the detection of $|{{Z}}_{k}|$ measurement cells can be approximately expressed as $\mathcal{O}\big( |{{Z}}_{k}|\xi\big)$. Let $\Xi\approx\text{P}_{fa}|{{Z}}_{k}|$ denote the number of point measurements after threshold detection with a constant $\text{P}_{fa}$. The tracker primarily utilizes $\Xi$ measurements to update target states or initiate a new target track, thus the complexity is around $\mathcal{O}\big(\Xi\big)$. The total computational complexity of DBT is around $\mathcal{O}\big(|{{Z}}_{k}|\xi+\Xi\big)$.

\emph{2) {\color{black}SPE-MF-TBD} Method:} It is shown in Fig.~\ref{fig: TBD} that the computational complexity of {\color{black}SPE-MF-TBD} is mainly from the multi-frame integration procedure. During each update of the merit-function, it needs to calculate the state transition set $\tau_\text{{\color{black}SPE}}({\color{black}\mathbf{x}}_{k-1})$ including the matrix inverse operations of $\bm{Q}_{k|k-1}$, and maximization operation $\max_{\tau_\text{SPE}^{-1}(\cdot)}$, according to (\ref{eq:grid_state_possible}) and (\ref{eq:KM_detect1_AMM1}), respectively. Let $\tau=|\tau_\text{{\color{black}SPE}}({\color{black}\mathbf{x}}_{k-1})|$, $k=1,\dots,K-1$, denote the number of elements in the state transition set for convenience. For convenience, the complexity of the operation $\bm{Q}_{k|k-1}^{-1}$ is denoted as $\mathcal{O}(\zeta)$, the complexity of the maximization operation with $\tau$ samples is $\mathcal{O}(\tau)$.
Each update of the merit-function $\mathcal{I}({{\color{black}\mathbf{x}}}_{k})$ requires the iteration of all states in sets ${\mathbb{G}_{k}^{4\times1}}$ and ${\mathbb{T}_{k-1}}$. Thus, it has a worst case complexity $\mathcal{O}\big(\tau(1+\zeta)(|{\mathbb{G}_{k}^{4\times1}}|+|{\mathbb{T}_{k-1}}|)\big)$. A complete batch 
processing needs to update merit-function $K-1$ times. Thus the computational complexity of multi-frame integration is $\mathcal{O}\big(\tau(1+\zeta)\sum_{k=2}^K(|{\mathbb{G}_{k}^{4\times1}}|+|{\mathbb{T}_{k-1}}|)\big)$.

\begin{table}
\setlength{\abovecaptionskip}{0.05cm} 
\centering
\caption{\textbf{Automotive Radar Parameters}} 
\begin{tabular*}{\hsize}{@{\extracolsep{\fill}}c c c c} 
  \toprule 
   Parameters& Values&Parameters& Values\\
  \hline 
  $\Delta_r$ & $0.5$~m & $[r_\text{min},r_\text{max}]$ & $[0.2,35]$~m \\
  $\Delta_d$ & $1.5$~m$/$s & $[d_\text{min},d_\text{max}]$ & $[-33.3,33.3]$~m$/$s \\
  $\Delta_{\theta}$ & $4.5^{\circ}$ & $[\theta_\text{min},\theta_\text{max}]$ & $[-45,45]^{\circ}$ \\
  $T$ & $70$~ms & $\beta$ & $-28^{\circ}$\\
  \toprule 
\end{tabular*}
\label{ta:sessdfsf1}
\end{table}

Note that the variable $\bm{Q}_{k|k-1}^{-1}$ can be calculated and stored before radar measurements. In fact, the radar FOV $[r_\text{min},r_\text{max}]$, $[d_\text{min},d_\text{max}]$,  $[\theta_\text{min},\theta_\text{max}]$, and the resolution cells $\Delta_r\times\Delta_d\times\Delta_{\theta}$ can be determined after the radar product is finalized. The covariance matrix $\bm{Q}_{k}^\text{p}$ of the self-positional error of the ego-vehicle can be tested before use. Other variables, such as $\bm{Q}_{k-1}^\text{t}$, $[\dot{\theta}_{\text{max}},\dot{\theta}_{\text{min}}]$, are artificially parameterized. In practical operations, we can calculate in advance of  $\bm{Q}_{k|k-1}^{-1}$ using historical multi-frame location information of the ego-vehicle. The computational complexity can be recast as $\mathcal{O}\big(\tau\sum_{k=2}^K(|{\mathbb{G}_{k}^{4\times1}}|+|{\mathbb{T}_{k-1}}|)\big)$. Generally, 
the merit-function of all elements in set ${\mathbb{G}_{k}^{4\times1}}$ or ${\mathbb{T}_{k-1}}$ at each scan can be updated in parallel. 
Thus, the above computational complexity can be further reduced by designing parallel processing architectures. In addition, some strategies including low-threshold preprocessing~\cite{grossi2013novel,10814330}, and recursive multi-frame integration~\cite{WangTAES} can be introduced to further reduce the computational complexity of the algorithm.
}

\begin{figure}
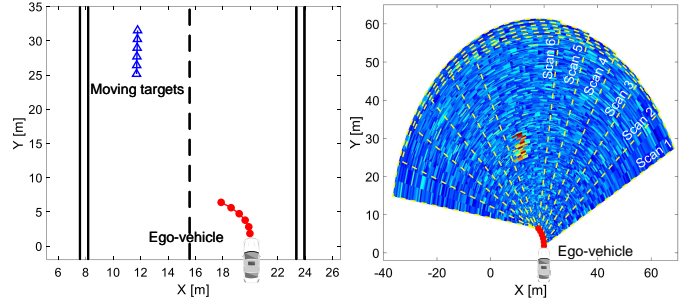

	\centering	\subfigure{\centering\includegraphics[width=1.72in]{CT_SC_v1.pdf}}
	\subfigure{\centering\includegraphics[width=1.72in]{CT_M.pdf}}	
    	\caption{(Left) An ego-vehicle with a CT motion, and (Right) its radar measurements of an overlaid snapshot of $6$ scans. Each scan corresponds to a fixed-size sector of the radar FOV. }
	\label{fig: ECHO_ct}
\end{figure}

\section{Simulation Results and Discussion}
\label{sec:Results}

\begin{figure}
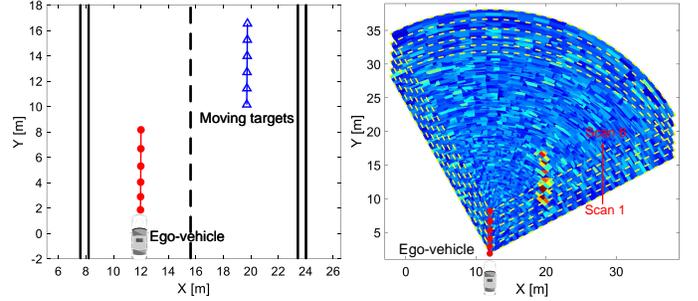

	\centering
	\subfigure{\centering\includegraphics[width=1.72in]{CA_SC_v1.pdf}}
	\subfigure{\centering\includegraphics[width=1.72in]{CA_M.pdf}}	
	\caption{ (Left) An ego-vehicle with a CA motion, and (Right) its radar measurements of an overlaid snapshot of six scans. }
	\label{fig: ECHO_cA}
\end{figure}
This section evaluates the performance of the proposed {\color{black}SPE-MF-TBD} 
approach for 
automotive radar scenarios by  
comparing it with {\color{black}the classical MF-TBD~\cite{grossi2013novel}, DBT, and G-TBD~\cite{Chen_sensors} methods. }

Here, the classical MF-TBD considering a stationary platform usually uses a rough velocity or acceleration upper bound of the target to integrate target energy among different frames~\cite{grossi2013novel,orlando2011track,WEIYI2019TVT}. Existing widely used detection and tracking algorithms on automotive radars first proceed with an SFD procedure to acquire point cloud measurements, then extended Kalman filter (EKF) for nonlinear conversion is implemented to estimate target  tracks~\cite{Kay,Bar2002Frontmatter}, which is referred to as DBT. {\color{black}G-TBD uses a two-stage detection architecture~\cite{Chen_sensors}. The first stage is the classical detector and plot-extractor to obtain a set of candidate plots. Then, a graph based radar target
detection algorithm is given to extract target track from multi-frame candidate plots.}
The following performance metrics are considered in this section: 1) The probability of detection ($P_{d}$) is defined as the probability that the test signal exceeds the detection threshold, and its position difference with the real target position is within two resolution cells. 2) The root mean square error (RMSE) of the declared target track satisfying the above detection conditions in the position dimension is given by  
\begin{align}	\text{RMSE}&=\sqrt{\sum_{n=1}^N\sum_{k=1}^K\frac{(x_{k,n}-\hat{x}_{k,n})^2+(y_{k,n}-\hat{y}_{k,n})^2}{NK}},\label{eq:RMSE1}
\end{align}
where $n$ is the $n$th realization of all $N$ Monte Carlo trials.  
The performance metrics $P_{d}$ and RMSE are evaluated by resorting to $500$ independent trials.

Consider a traffic target detection scenario, the echo measurements are collected from a FMCW radar system mounted on a moving ego-vehicle.
The centroid state ${\color{black}\mathbf{p}}_k$ of the ego-vehicle can be measured through the navigation system with a measurement error $\bm{\omega}^\text{p}_{k}$. The automotive radar parameters are given in Table~\ref{ta:sessdfsf1}.
Some typical movement scenarios of the ego-vehicle, such as CA and CT models,  are considered to evaluate the performance of the proposed algorithms. Considering target motion in practical traffic scenarios, the maximum acceleration and yaw rate of a moving target are $a_{\text{max}}=33~$m$/\text{s}^2$ and $\dot{\alpha}_{\text{max}}=50~^{\circ}/\text{s}$, respectively.
The SNR of the target is defined as $\text{SNR}=10\log(A^2/\sigma^2)$.
\begin{figure}
	\centering	\subfigure{\centering\includegraphics[width=1.735in]{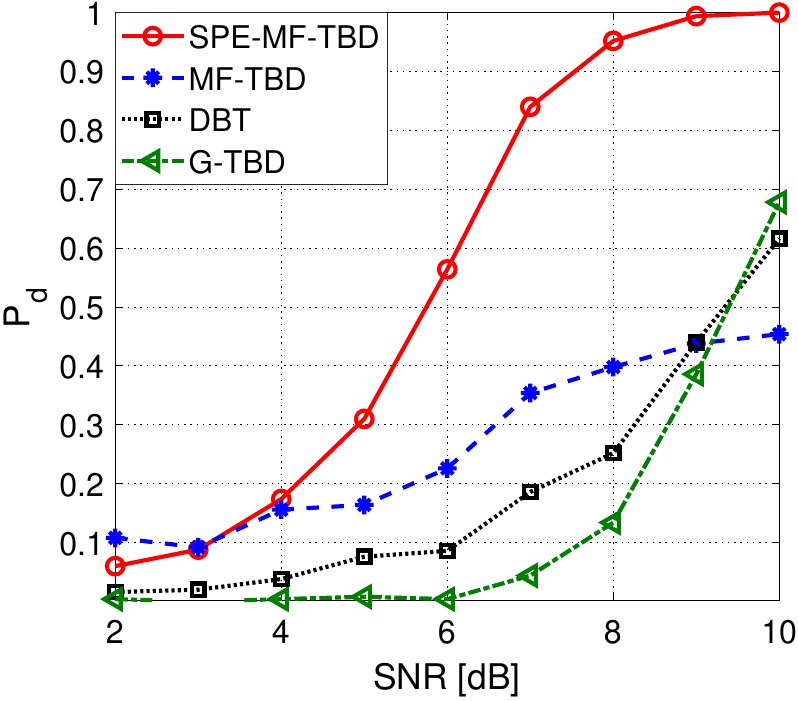}}
	\subfigure{\centering\includegraphics[width=1.705in]{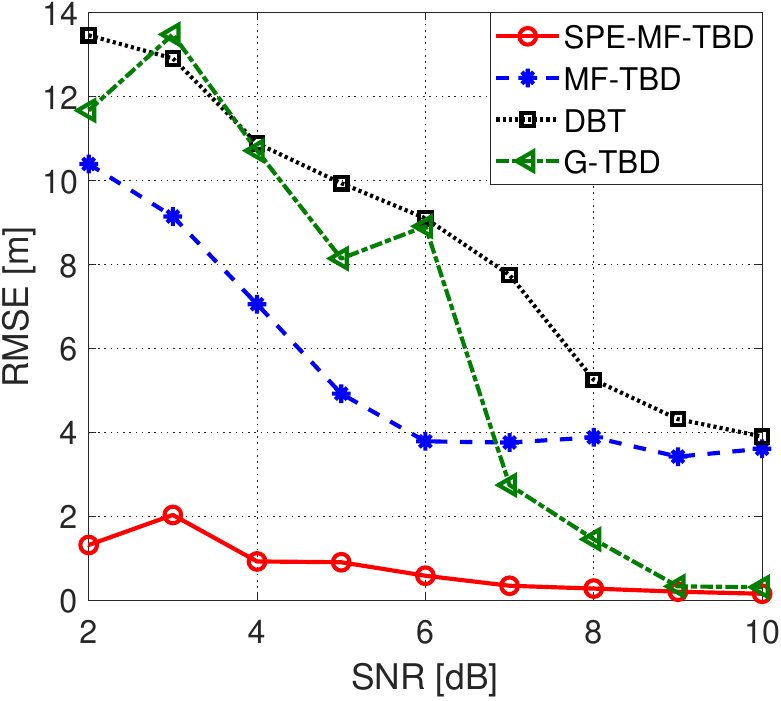}}	
	\caption{ (Left) $P_{d}$ and (Right) RMSE versus SNR for the CT motion ego-vehicle when $\eta=10$, and $K=6$.  }
	\label{fig: PD_ct}
\end{figure}

According to~\cite{Sun_PHD,RN476}, the self-positioning errors of the ego-vehicle in the position $(p_{x,k},p_{y,k})$, velocity $(p_{\dot{x},k},p_{\dot{y},k})$ 
and yaw angle $\alpha_k$ dimensions are around $(e_{x,k},e_{y,k})=(1.69~\text{m},0.83~\text{m})$, $(e_{\dot{x},k},e_{\dot{y},k})=(0.04~\text{m}/\text{s},0.04~\text{m}/\text{s})$ and $e_{\alpha,k}=2.54~^{\circ}$, respectively. Then, the covariance matrix of self-positioning errors in (\ref{eq:Vehicle_STATE}) can be approximately expressed as 
\begin{equation}
\bm{Q}_{k}^\text{p}=\text{diag}(\,e_{x,k}^2,e_{\dot{x},k}^2,e_{y,k}^2,e_{\dot{y},k}^2,e_{\alpha,k}^2)/(3\eta),
\end{equation}
where $\eta=1,2,4,6,8,10$ is a preset error factor. Note that the error covariance matrix decreases with increasing $\eta$. 
In the following analysis, the detection thresholds of different algorithms are set based on a constant $P_{fa}=1\times10^{-3}$. 
The above parameter values of radar systems, moving targets, and ego-vehicles can be set arbitrarily, and they do not affect the generality of algorithms.


\begin{figure}
	\centering
\subfigure{\centering\includegraphics[width=1.735in]{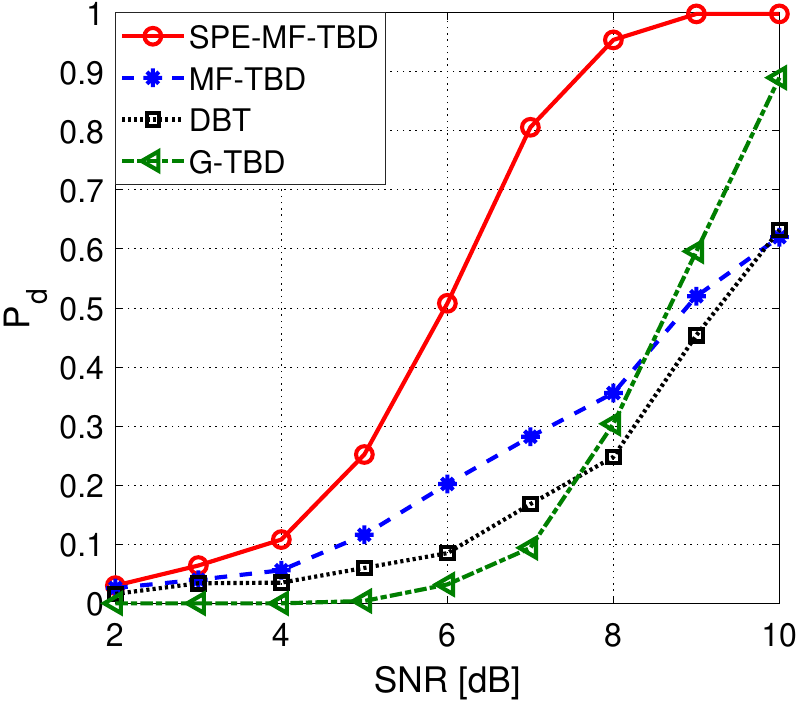}}	\subfigure{\centering\includegraphics[width=1.705in]{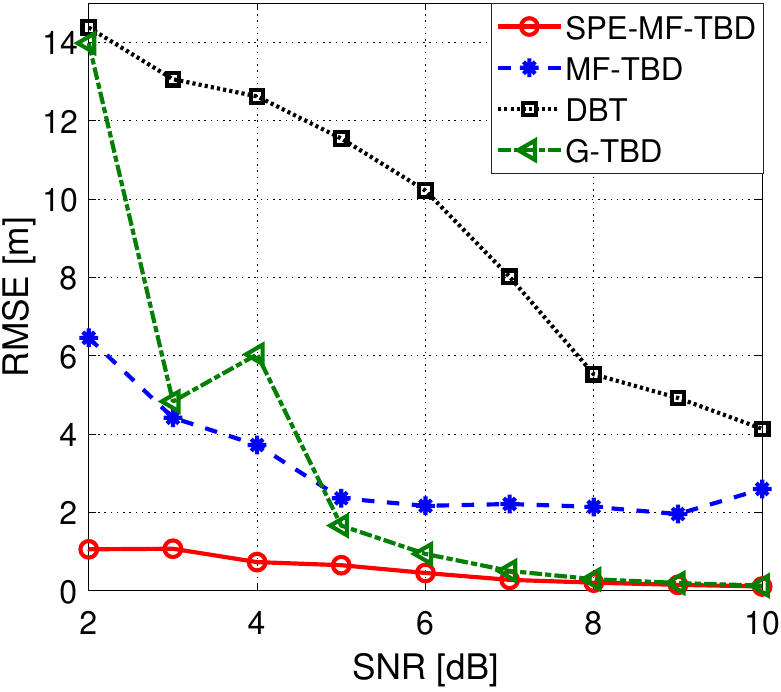}}		\caption{(Left) $P_{d}$ and (Right) RMSE versus SNR for the CA motion ego-vehicle when $\eta=10$, and $K=6$. }
	\label{fig: PD_ca}
\end{figure}

\subsection{Scenario~1}
\label{subsec: Scenario_1}

To verify the robustness of the proposed {\color{black}SPE-MF-TBD} algorithm, two classical motion models, such as CT and CA models with unknown acceleration and turn rate, respectively, for the ego-vehicle are considered. 
Here, the values of acceleration $a$ and the turn rate $\varphi$ are uniformly generated at each Monte Carlo with $a\in [0, 28]~\text{m}/\text{s}^2$ and $\varphi\in [0, 0.873\pi]~\text{rad}/\text{s}$, respectively. In this scenario, we assume that the target will remain within the radar FOV during $K$ scans. 
\begin{figure}
	\centering	\subfigure{\centering\includegraphics[width=1.735in]{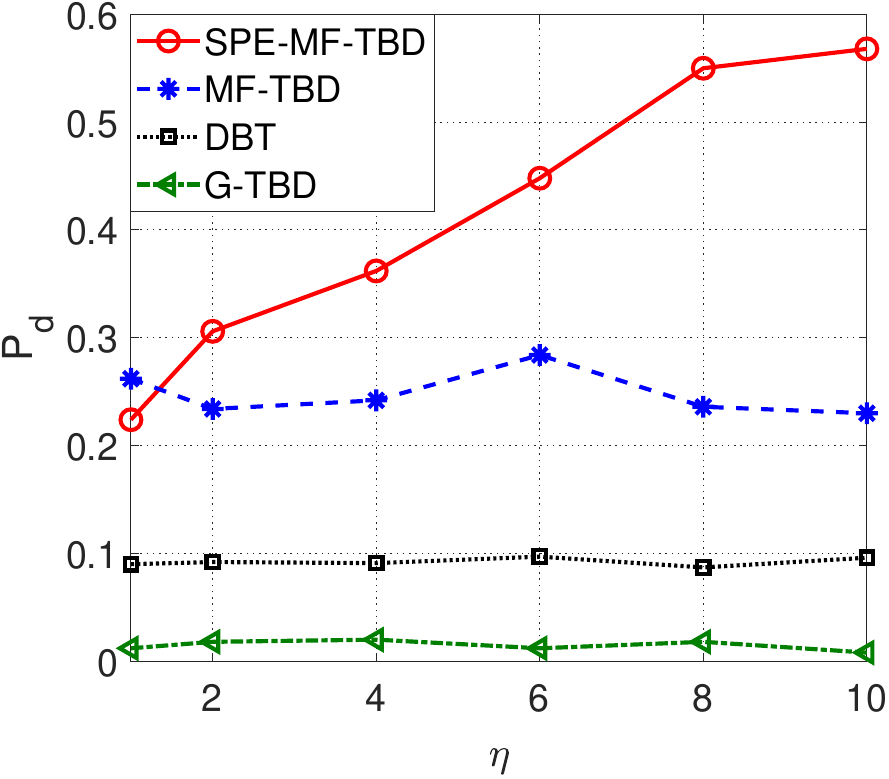}}	\subfigure{\centering\includegraphics[width=1.705in]{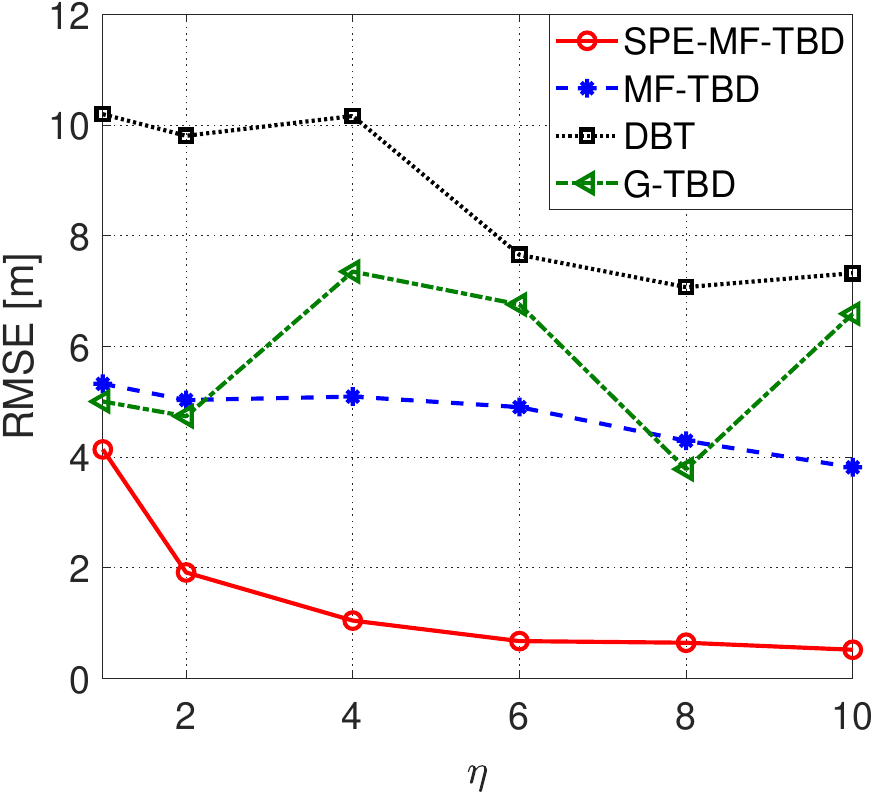}}
	\caption{{\color{black}(Left) $P_{d}$ and (Right) RMSE versus the positioning error factor $\eta$ for the CT motion ego-vehicle when SNR~$=6$~dB, and $K=6$. }}
	\label{fig: PD_CT}
\end{figure}
\begin{figure}
	\centering	\subfigure{\centering\includegraphics[width=1.735in]{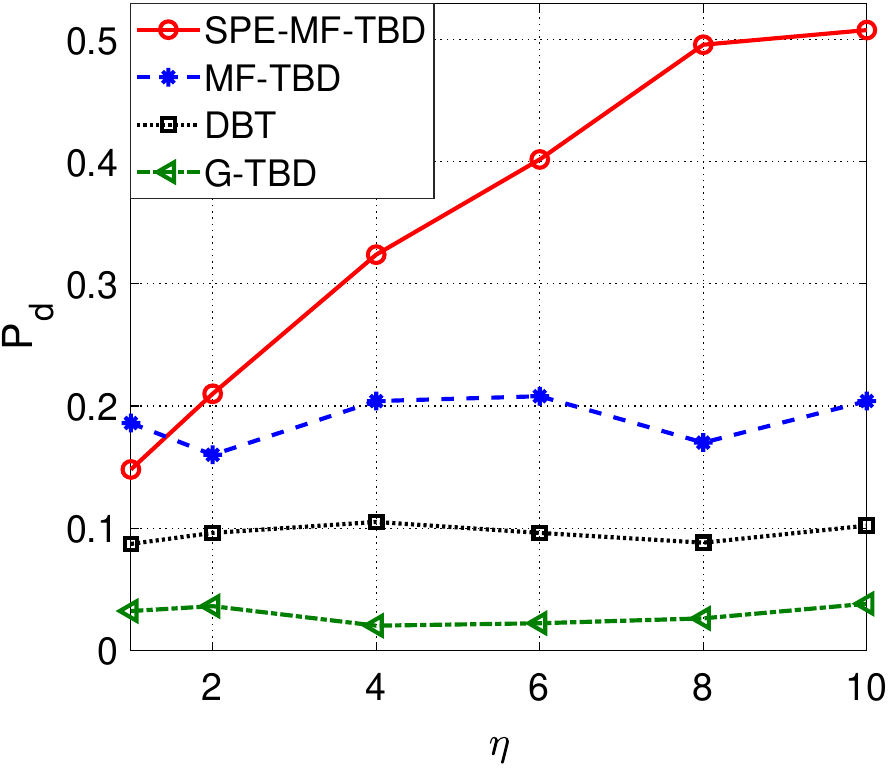}}	\subfigure{\centering\includegraphics[width=1.705in]{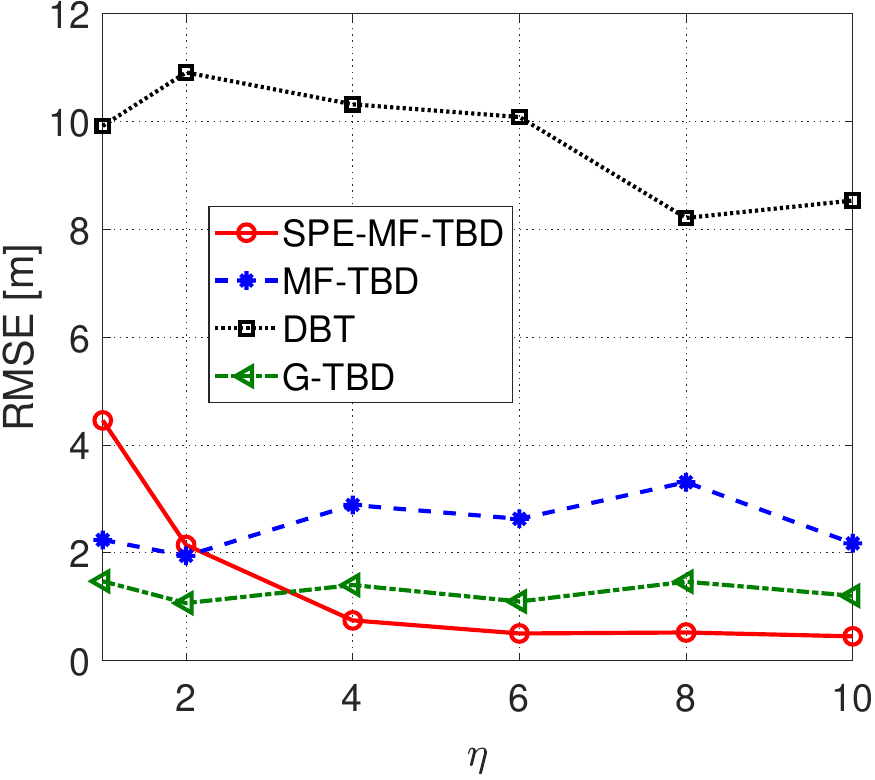}}	
	\caption{{\color{black}(Left) $P_{d}$ and (Right) RMSE versus the positioning error factor $\eta$ for the CA motion ego-vehicle when SNR~$=6$~dB, and $K=6$.} }
	\label{fig: PD_CA}
\end{figure}

The echo measurements of the automotive radar when the ego-vehicle satisfies CT and CA motion models are shown in Figs.~\ref{fig: ECHO_ct} and \ref{fig: ECHO_cA}, respectively. It can be seen that the radar FOV changes with ego-vehicle movements, such as sudden turns and accelerations.

\begin{figure}
	\centering	\subfigure{\centering\includegraphics[width=1.735in]{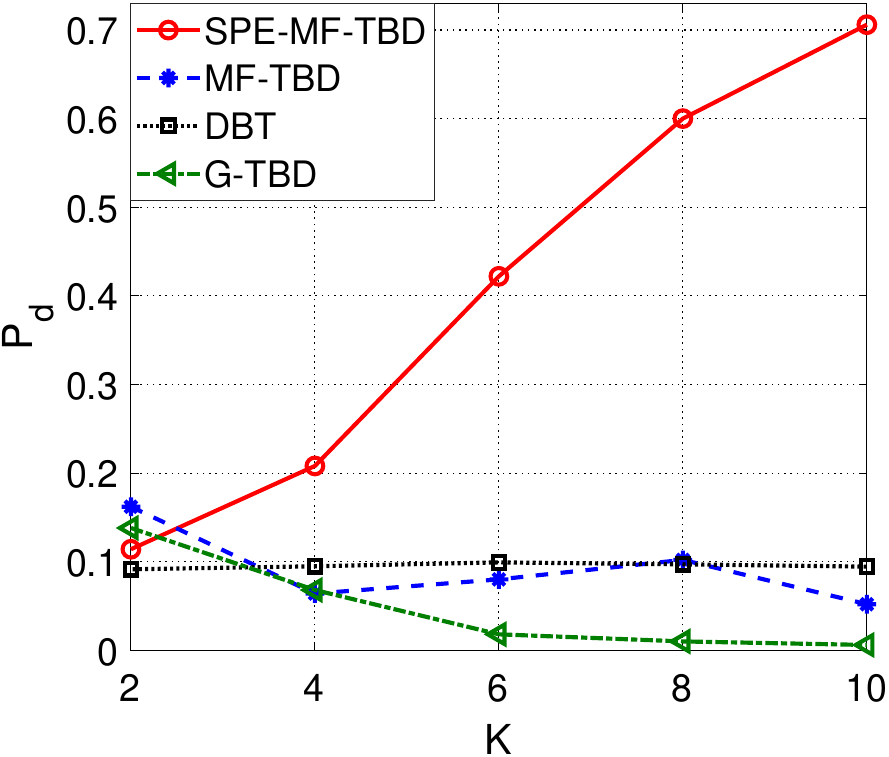}}	\subfigure{\centering\includegraphics[width=1.705in]{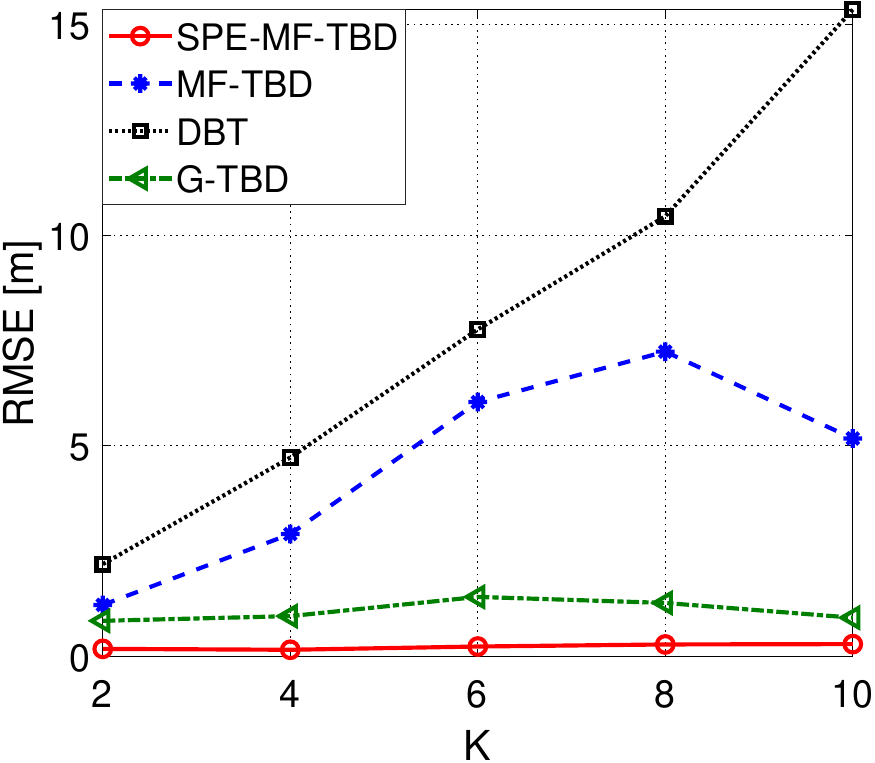}}	
	\caption{{\color{black}(Left) $P_{d}$ and (Right) RMSE versus the number $K$ of frames accumulated for the CT motion ego-vehicle when SNR~$=6$~dB, and $\eta=10$.} }
	\label{fig: PD_K_CT}
\end{figure}

In Figs.~\ref{fig: PD_ct} and \ref{fig: PD_ca}, we investigate the detection and tracking
performance of algorithms in the CT and CA motion scenarios for an ego-vehicle when $\eta=10$, and $K=6$, respectively.  It turns out that, for different values of the target SNR, the proposed {\color{black}SPE-MF-TBD} algorithm also outperforms MF-TBD, G-TBD and DBT methods in terms of detection probability and tracking accuracy. Traditional MF-TBD algorithms built on a stationary radar platform suffer from severe performance degradation in the detection and tracking by ignoring the movement of the ego-vehicle. MF-TBD  still maintains slightly better detection and tracking performance than DBT at low target SNR, but this performance gap gradually decreases as the SNR increases, and it even occurs that DBT outperforms MF-TBD
at a higher SNR. {\color{black}G-TBD is a two-stage detection architecture that experiences information loss after single-frame detection. Therefore, G-TBD exhibits lower detection probability and tracking accuracy than SPE-MF-TBD for low SNR targets, but it can achieve tracking accuracy comparable to SPE-MF-TBD when the SNR is sufficiently high. In addition, G-TBD requires the yaw angle of the ego vehicle to be known and constant\cite{Chen_sensors}. As a result, its degradation in detection and tracking performance for the CT motion ego-vehicle can be observed by comparing Fig.~\ref{fig: PD_ct} and Fig.~\ref{fig: PD_ca}.
}

\begin{figure}[!]
	\centering	\subfigure{\centering\includegraphics[width=1.735in]{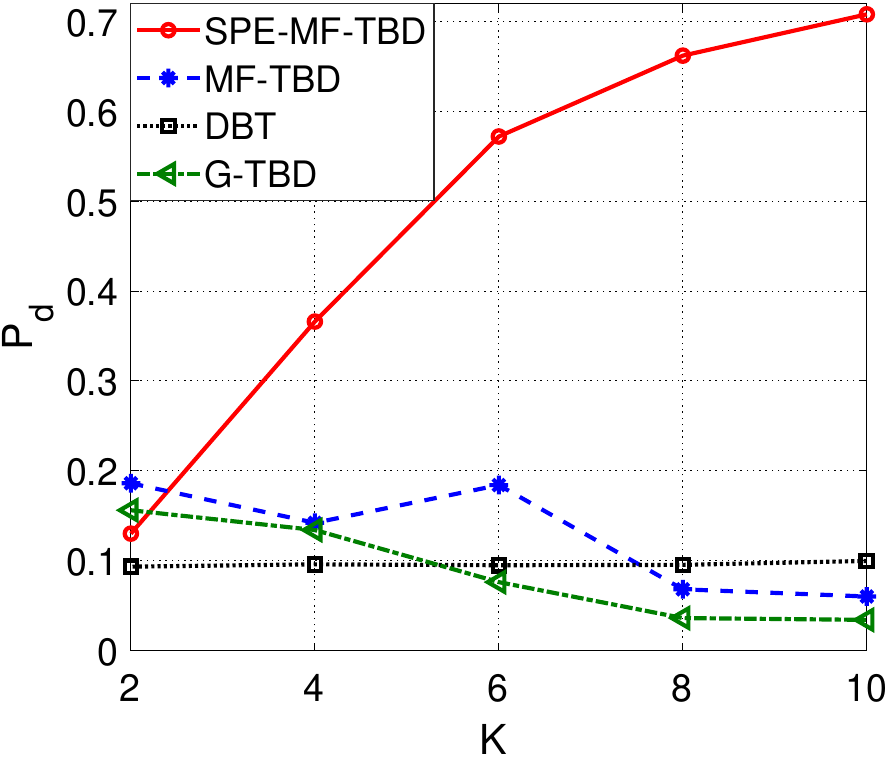}}	\subfigure{\centering\includegraphics[width=1.705in]{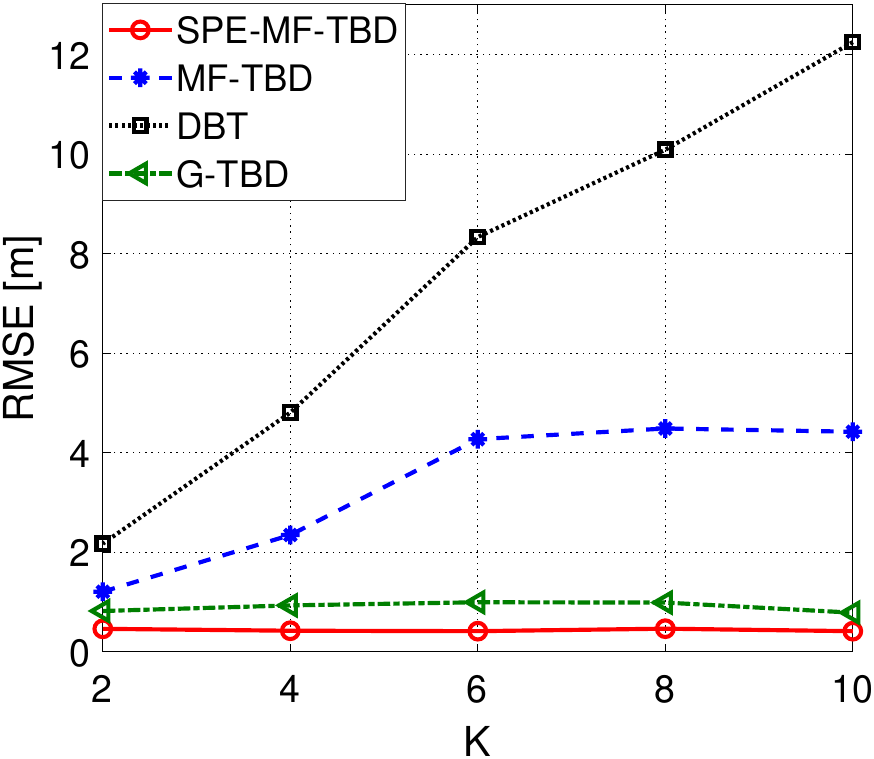}}	
	\caption{{\color{black}(Left) $P_{d}$ and (Right) RMSE versus the number $K$ of frames accumulated for the CA motion ego-vehicle when SNR~$=6$~dB, and $\eta=10$.}} 
	\label{fig: PD_K_CA}
\end{figure}
 

\begin{table}
\setlength{\abovecaptionskip}{0.05cm} 
\centering
\caption{\textbf{{\color{black}Average Running Times of Algorithms (Unit: [s])}}} 
{\color{black}
\begin{tabular*}{\hsize}{@{\extracolsep{\fill}}c c c c c c} 
  \toprule 
   $K$ & $2$& $4$ & $6$ & $8$ & $10$\\
  \hline 
  SPE-MF-TBD & $1.89$ & $5.98$ & $11.27$ &$18.23$&$25.91$ \\
   \hline
  G-TBD & $0.09$ & $0.27$ & $0.53$ &$0.74$& $0.95$\\
  \hline
  MF-TBD & $0.06$ & $0.12$ &  $0.18$&$0.25$&$0.39$\\
  \hline
  DBT & $0.004$ & $0.007$ & $0.008$&$0.011$& $0.013$\\
  \toprule 
\end{tabular*}}
\label{ta:time}
\end{table}

The $P_{d}$ and RMSE of {\color{black}SPE-MF-TBD}, MF-TBD, {\color{black}G-TBD}, and DBT methods for the CA and CT motion scenarios of the ego-vehicle are plotted versus $\eta$ when SNR~$=6$~dB and $K=6$ in Figs.~\ref{fig: PD_CT} and \ref{fig: PD_CA}, respectively. 
It can be seen that the detection probability and tracking accuracy of the {\color{black}SPE-MF-TBD} algorithm increase as $\eta$ increases (or the decrease of self-positioning error covariance $\bm{Q}_{k}^\text{p}$ of the ego-vehicle) due to the consideration of position, velocity, and yaw angle errors of the ego-vehicle. 
The $P_d$ and RMSE of {\color{black}G-TBD}, MF-TBD and DBT approaches remain essentially unchanged for different values of $\eta$, since these methods ignore positioning errors of the ego-vehicle and directly assume that the navigation system is capable of providing real position, velocity, and yaw angle information.
However, the proposed  {\color{black}SPE-MF-TBD} algorithm still shows better detection and tracking performance than the traditional MF-TBD, {\color{black}G-TBD} and DBT approaches in weak target scenarios. 
Here, the detection and tracking performance of MF-TBD exceeds DBT. {\color{black}As expected, G-TBD exhibits low detection probability and tracking accuracy due to model mismatch, particularly in CT motion scenarios.

In Figs.~\ref{fig: PD_K_CT} and \ref{fig: PD_K_CA}, in the case of $\text{SNR}=6$~dB, $\eta=10$, the curves of $P_{d}$ and RMSE for CA and CT motion scenarios of
the ego-vehicle are presented based on the different number of frames accumulated, $K=2,4,6,8,10$. It can be seen that the detection probability of SPE-MF-TBD increases as the increase of $K$. The tracking accuracy of SPE-MF-TBD remains almost constant and consistently outperforms the traditional MF-TBD, G-TBD, and DBT approaches for different values of $K$ when $\text{SNR}=6$~dB.
The $P_{d}$ of DBT is independent to $K$ due to the single-frame detection mechanism. MF-TBD and G-TBD  struggle to maintain improved detection performance as the number of accumulated frames increases, primarily due to model mismatch. Furthermore, uncertainty in target motion grows with longer time intervals. }

{\color{black}In addition, the average running times of {\color{black}SPE-MF-TBD}, G-TBD, MF-TBD, and DBT algorithms are given in Table~\ref{ta:time}. It can be seen that the running time of SPE-MF-TBD is longer than that of G-TBD, MF-TBD, and DBT approaches. Furthermore, the average running time of all algorithms gets longer as $K$ increases. Note that some parallel processing architectures can be designed to reduce the computational complexity of SPE-MF-TBD, as indicated in Section~\ref{se:Complexity}.}
\subsection{Scenario~2}
\label{subsec: Scenario_2}
\begin{figure}
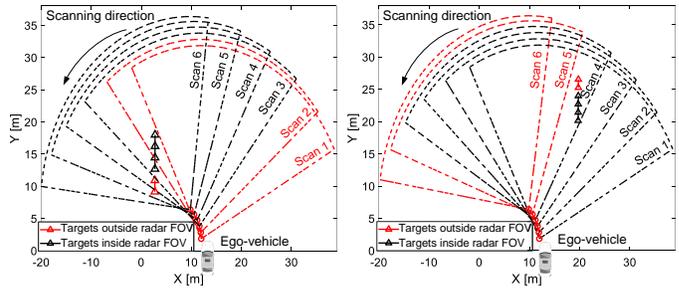

	\centering
	\subfigure{\centering\includegraphics[width=1.72in]{CT_outside_FOV2_Part_I.pdf}}
	\subfigure{\centering\includegraphics[width=1.72in]{CT_outside_FOV1_Part_I.pdf}}
	\caption{Target appears  (left subplot) and disappears (right subplot) from radar
FOV during $6$ scans when $\kappa=2$.  }
	\label{fig: LIMIT_FOV}
\end{figure}

In this experiment, we consider a more complicated case where the target appears and disappears from radar FOV during a batch processing of $K=6$ consecutive frames, as shown in Fig.~\ref{fig: LIMIT_FOV}. Specifically, targets fall into the outside of radar FOV during the first $\kappa$ scans and the second $\kappa$ scans with $1\leqslant\kappa\leqslant K$. Here, we mainly consider two cases with $\kappa=2,4$ for convenience, and 
two instances of $\kappa=2$ for a target that appears and disappears from radar FOV are shown in Fig.~\ref{fig: LIMIT_FOV}.

 \begin{figure}
	\centering
	\subfigure{\centering\includegraphics[width=1.73in]{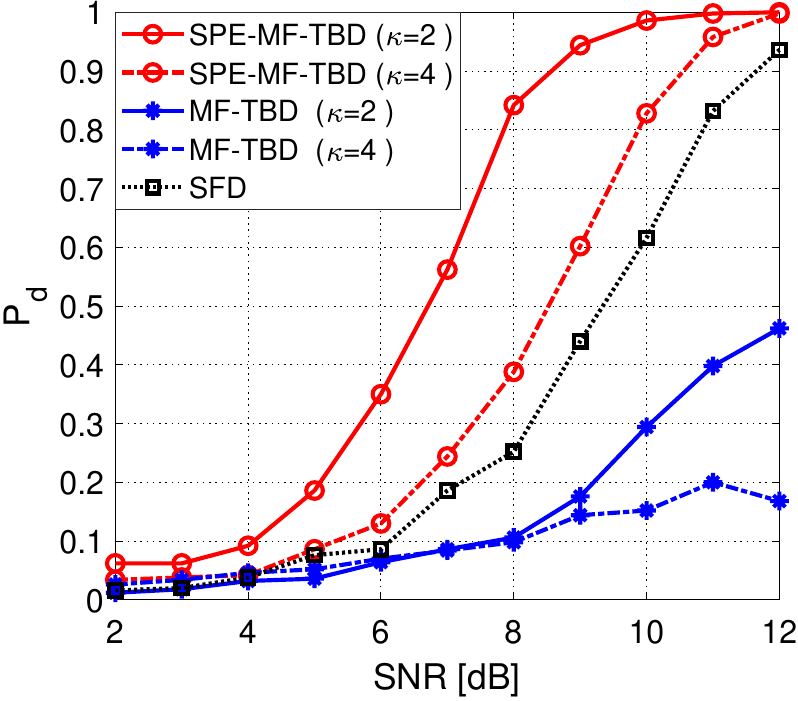}}	\subfigure{\centering\includegraphics[width=1.71in]{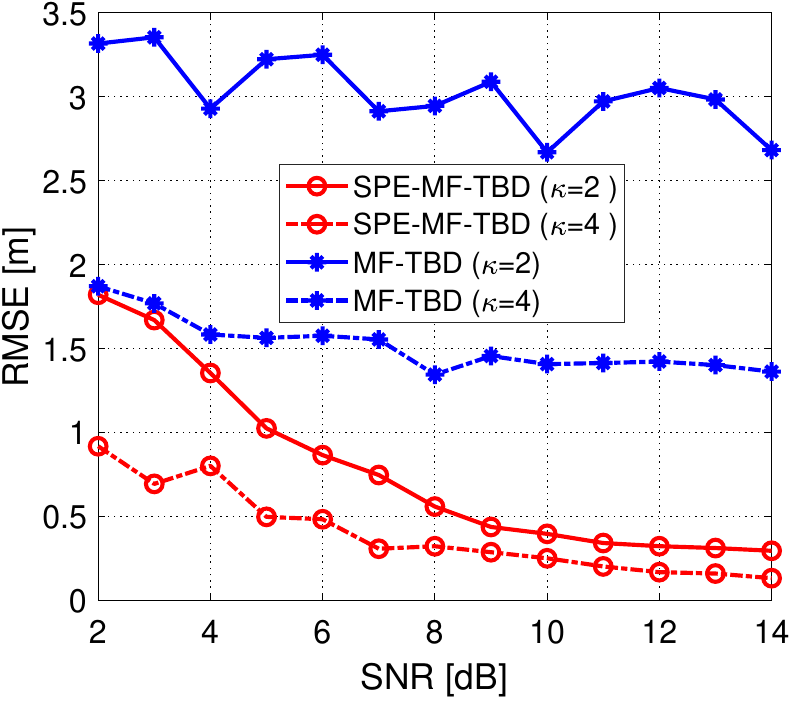}}
	\caption{(Left) $P_{d}$ and (Right) RMSE versus SNR when the target suddenly appears with $\kappa=2,4$, $\eta=10$, and $K=6$. }
	\label{fig: CM_birth}
\end{figure}

\begin{figure}
	\centering	\subfigure{\centering\includegraphics[width=1.73in]{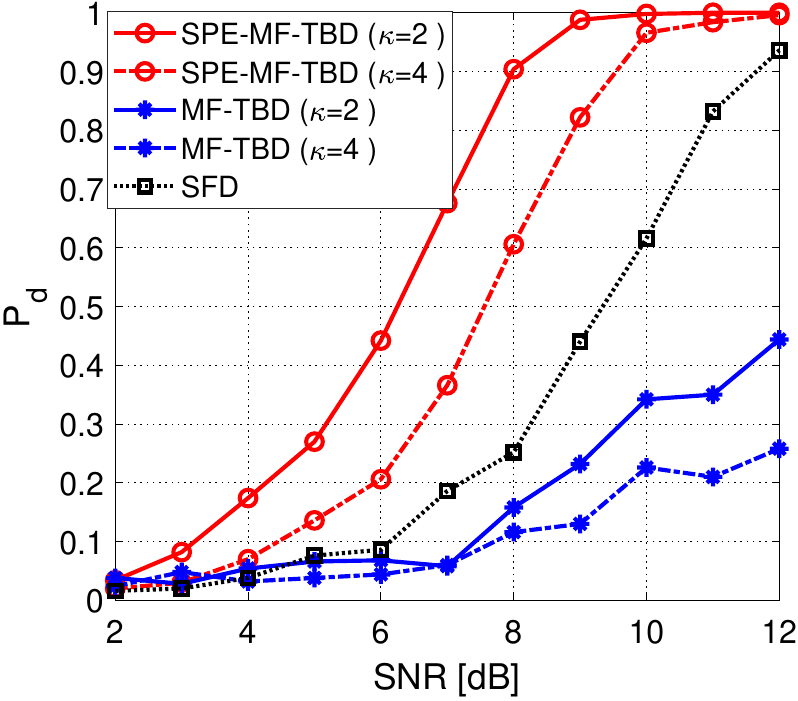}}	\subfigure{\centering\includegraphics[width=1.71in]{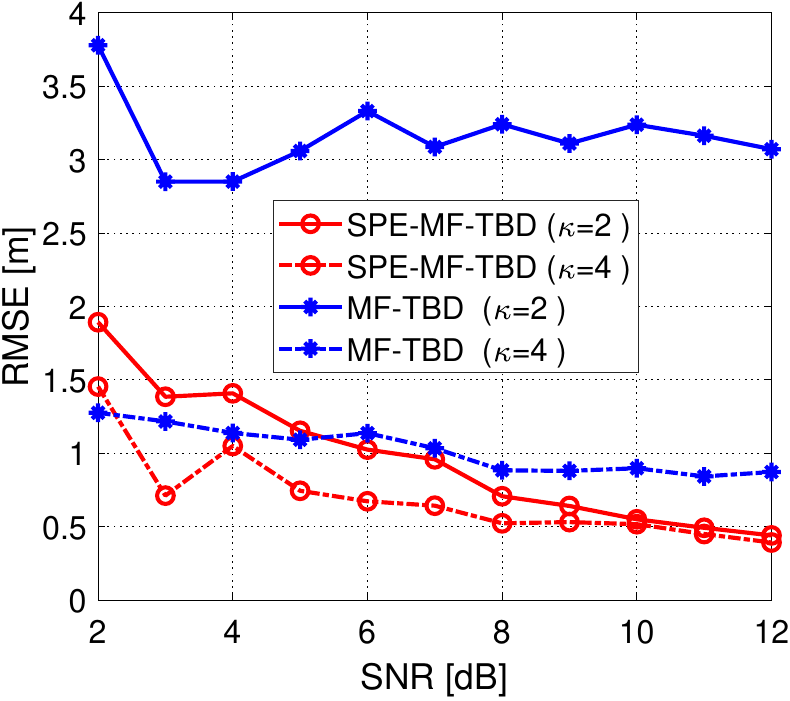}}
	\caption{(Left) $P_{d}$ and (Right) RMSE versus SNR when the target suddenly disappears with $\kappa=2,4$, $\eta=10$, and $K=6$. }
	\label{fig: CM_death}
\end{figure}

For the above scenarios, the proposed {\color{black}SPE-MF-TBD} algorithm can adaptively adjust the detection threshold according to the integrated path of (\ref{eq:satte})--(\ref{eq:KM_detect1_AMM_out}). {\color{black}
However, G-TBD assumes that the target remains in radar FOV during multiple scans, thus this method is not suitable in the case where targets appear or disappear during multi-frame integration. In this subsection, we mainly analyze the performance of SPE-MF-TBD and MF-TBD.}
Following that, the detection and tracking performances of {\color{black}SPE-MF-TBD}, and  MF-TBD algorithms when the target suddenly appears at times $k=3$ and $k=5$ with $K=6$ and $\eta=10$, are given in Fig.~\ref{fig: CM_birth}, where the $P_d$ curve of SFD is presented as a benchmark. Similarly, the $P_d$ and RMSE curves versus SNR when the target suddenly
disappears at times $k=5$ and $k=3$ with $K=6$ and $\eta=10$, are given in Fig.~\ref{fig: CM_death}. 
It can be seen that the $P_d$ and RMSE of the proposed  {\color{black}SPE-MF-TBD} approach are far superior to that of  MF-TBD algorithms for both scenarios with $\kappa=2$ and $\kappa=4$. As expected, the detection probability of {\color{black}SPE-MF-TBD} and MF-TBD decreases with increasing $\kappa$ due to the loss of more target information. However, the tracking accuracy of {\color{black}SPE-MF-TBD} and MF-TBD increases with increasing $\kappa$, since the number of statistical error variables is reduced with $\kappa$ increases according to the definition of RMSE  in (\ref{eq:RMSE1}).
Note that there is a serious deterioration in the detection performance of the traditional MF-TBD approach, which is even worse than that of DBT methods.  
The main reason is that MF-TBD suffers from a mismatch error during multi-frame integration due to the platform's motion. Besides, MF-TBD ignores the fact that the target falls outside the radar FOV, which further results in the inclusion of an additional noise signal in the integrated merit function of MF-TBD. Thus, the detection performance of MF-TBD will be further degraded, and its detection probability is difficult to reach 1, even at a high SNR.


 \begin{figure}
	\centering
\subfigure{\centering\includegraphics[width=3.65in]{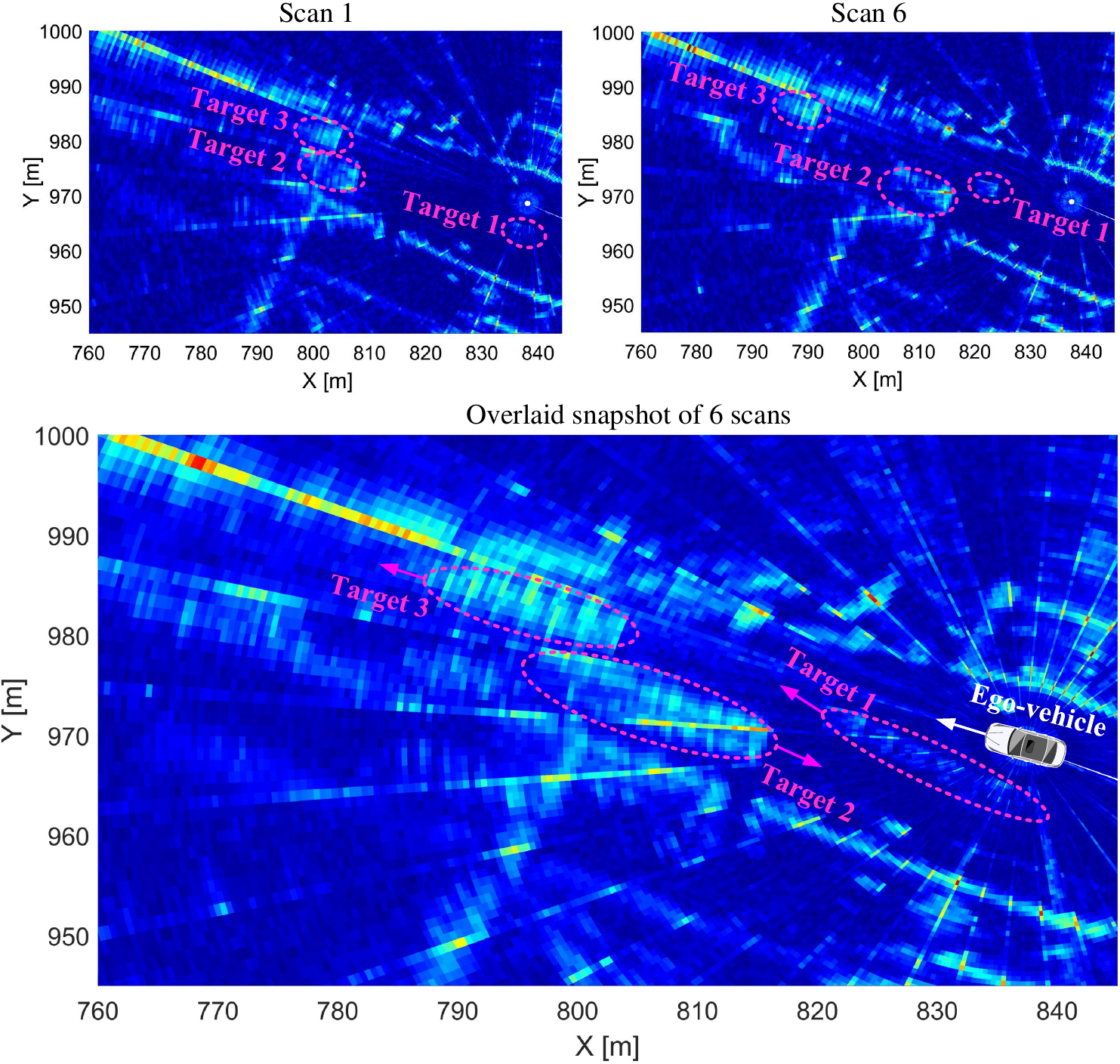}}
	\caption{Automotive radar echo measurements with scan~1, scan~6, and  an overlaid snapshot of $6$ scans.  }
	\label{fig: MEA}
\end{figure}

By comparing the $P_d$ curves in Figs.~\ref{fig: CM_birth} and ~\ref{fig: CM_death}, it is interesting to observe that 
a target that suddenly disappears from the radar FOV during multiple scans is more likely to be detected by the {\color{black}SPE-MF-TBD} algorithm and continue to be tracked for a while than a target that suddenly appears. Specifically, $P_d$ values of  {\color{black}SPE-MF-TBD} in  Fig.~\ref{fig: CM_death} are slightly larger than that in Fig.~\ref{fig: CM_birth}. In fact, the {\color{black}SPE-MF-TBD} algorithm sequentially integrates multi-frame echo measurements according to the radar scanning time sequence. For a target that suddenly appears during multi-frame integration,  {\color{black}SPE-MF-TBD} fails to establish the target confidence during the first few frames of integration and may introduce noise and interference due to missing targets. In this case, {\color{black}SPE-MF-TBD} has a slight loss of detection performance.

\section{Experimental Results With 
 Real Data}\label{sec:Scenario_Experimental}
This section evaluates the performance of the proposed  {\color{black}SPE-MF-TBD} algorithm using real measurements in the  Oxford Radar RobotCar Dataset~\cite{RadarRobotCarDatasetICRA2020}. This dataset is recorded from a $77$~GHz millimeter-wave scanning radar
with the range resolution of $4.38$~cm, and azimuth resolution of $0.9^{\circ}$ in urban environments. The radar is mounted at the top of the ego-vehicle, and its coverage is  $160$~m and $360^{\circ}$, with scan period $T=0.25$~s. The ego-vehicle accelerates along the lane with the maximum speed being up to $10$~m$/$s.  Automotive radar echo measurements with an overlaid snapshot of the first $K=6$ scans out of a total of $17$ scans are shown in Fig.~\ref{fig: MEA}. Here, the single-frame echo measurements, such as scan~1 and scan~6, are given for convenience. 
It is shown that there are $3$ targets moving in the radar FOV, and the echo energy of target 1 is relatively lower.
 \begin{figure}
	\centering
	\subfigure{\centering\includegraphics[width=1.72in]{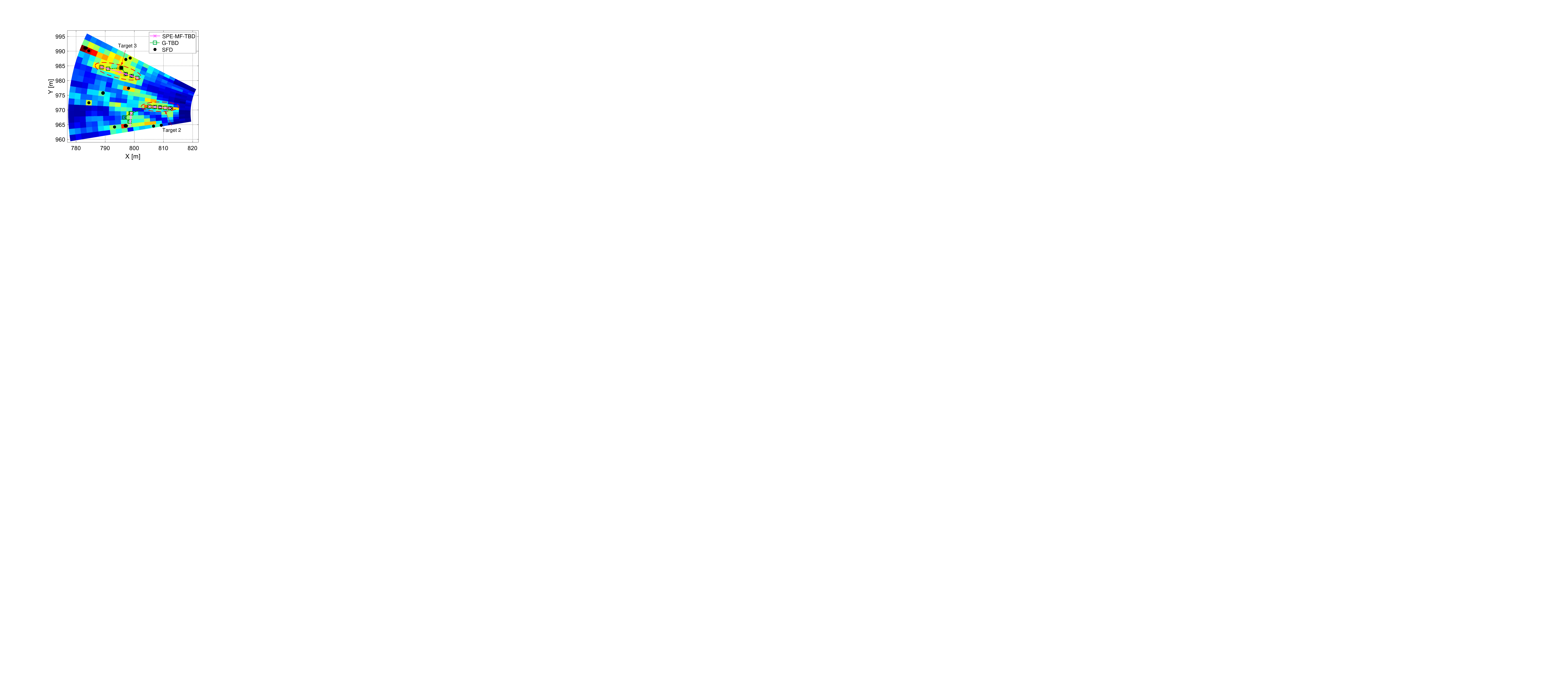}}	\subfigure{\centering\includegraphics[width=1.72in]{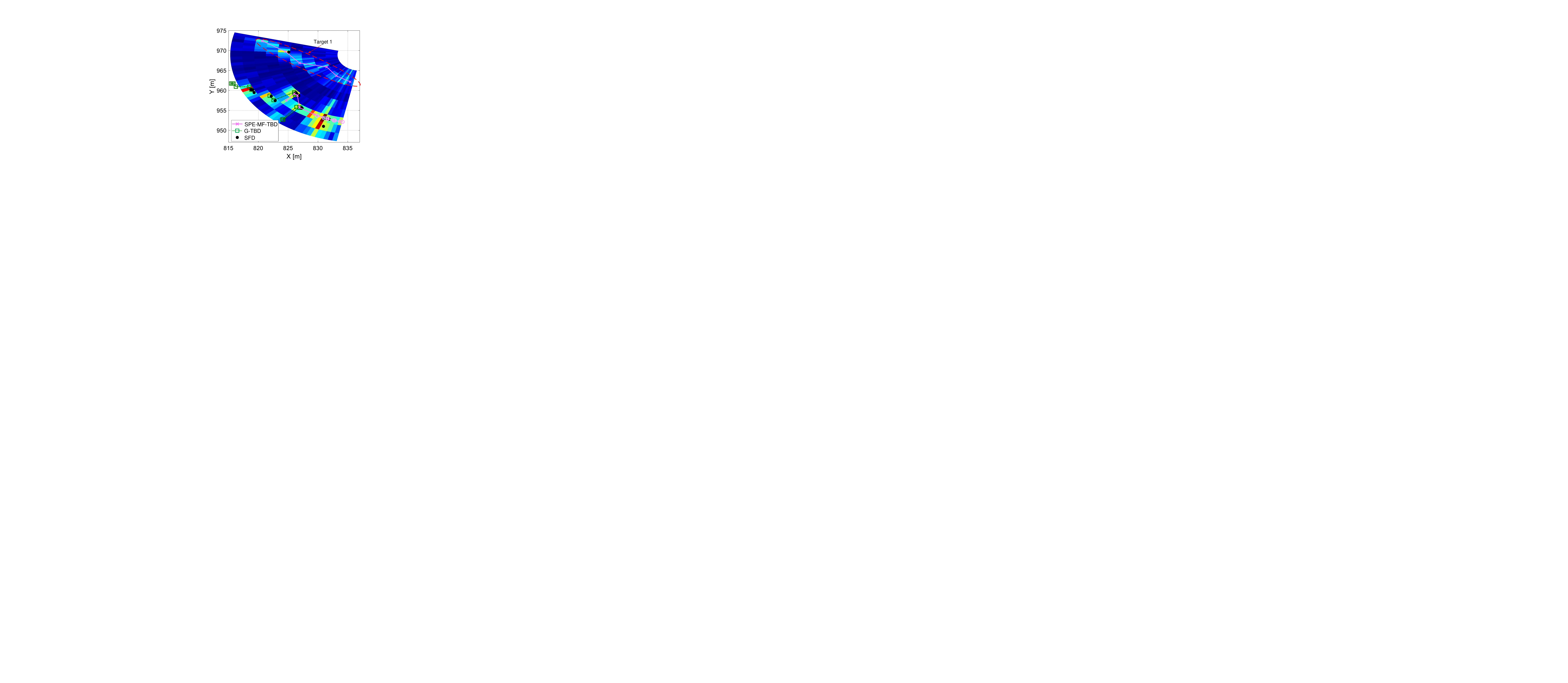}}
	\caption{Output tracks produced by {\color{black}SPE-MF-TBD and G-TBD}  at different times. The black point measurement from SFD is given as a benchmark. The background image is radar echo measurements with an overlaid snapshot of $6$ scans. 
 }
	\label{fig: zone_track}
\end{figure}

\begin{figure*}
	\centering	\subfigure{\centering\includegraphics[width=3.1in]{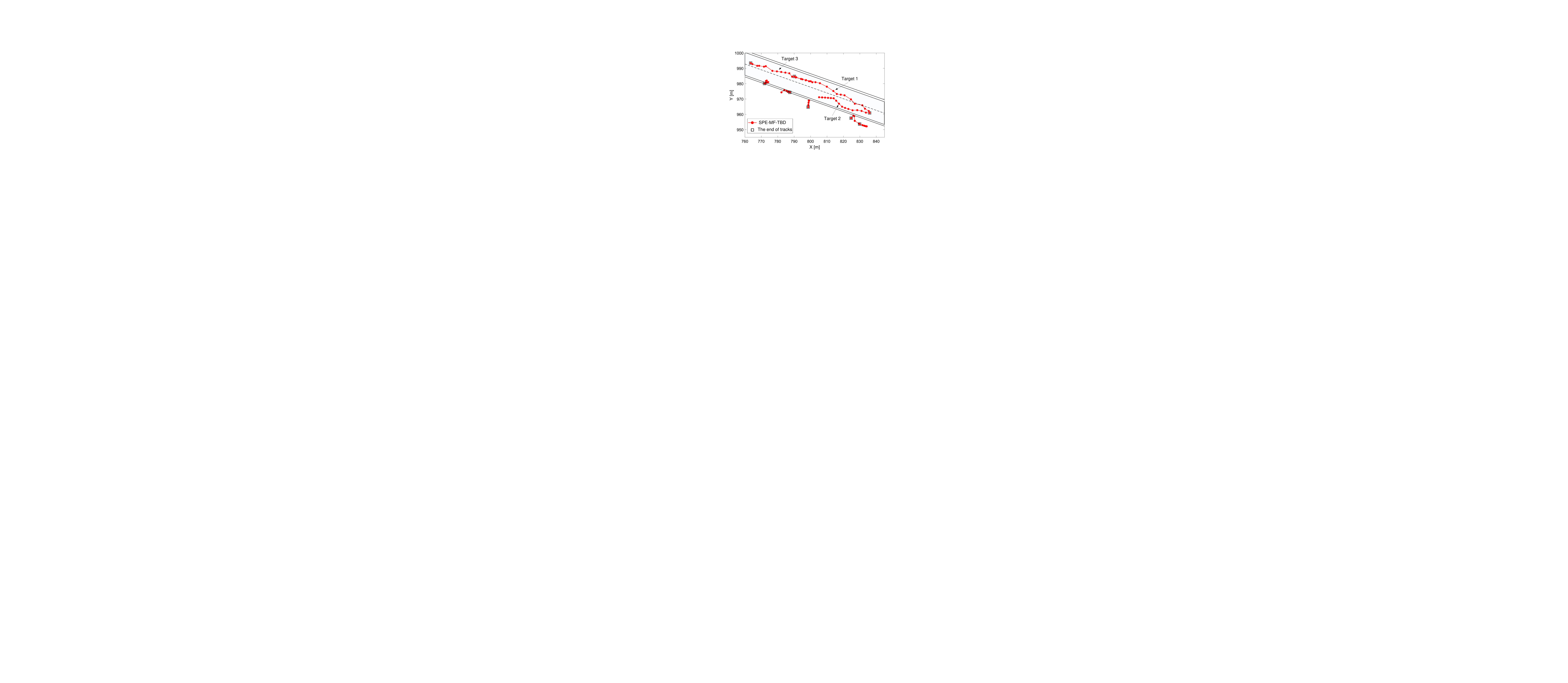}}
\hspace{12mm}
 \subfigure {\centering\includegraphics[width=3.1in]{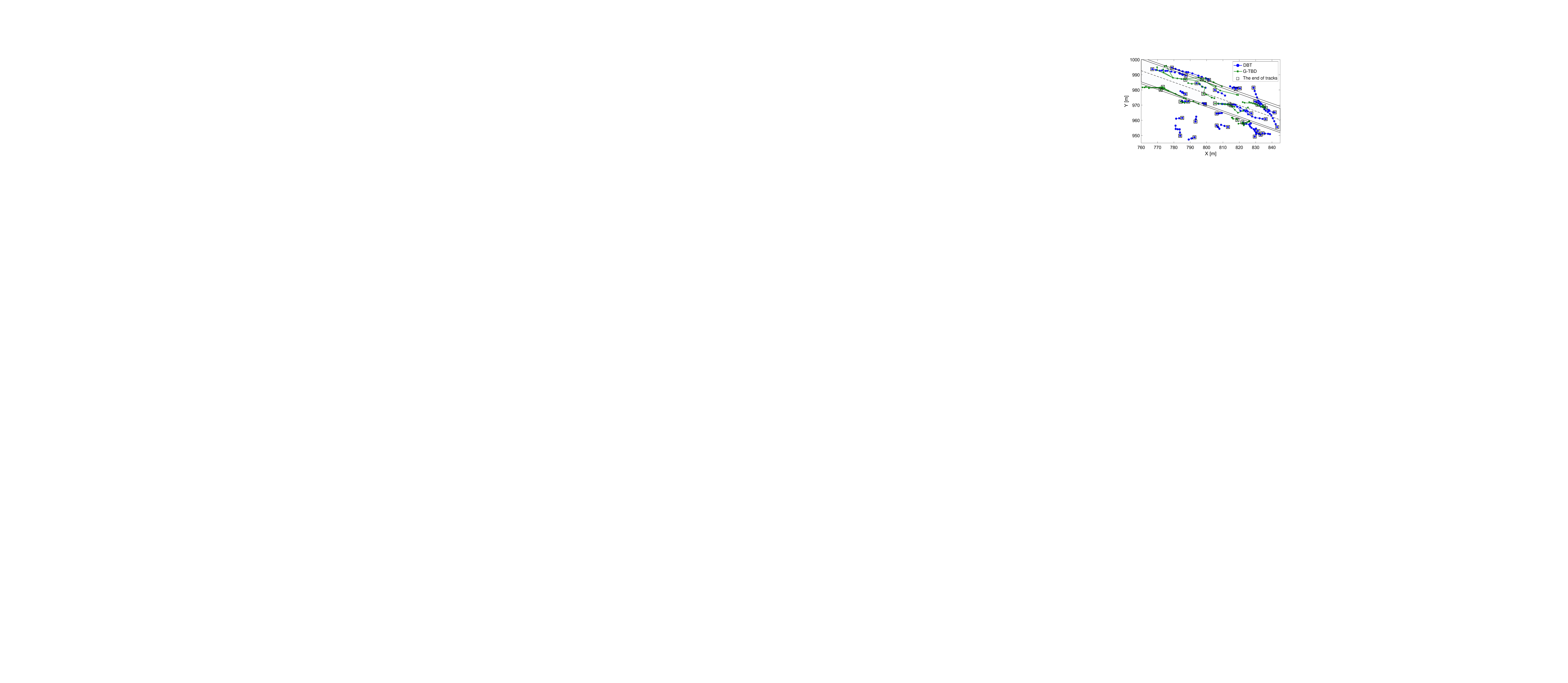}}
	\caption{Estimated tracks produced by {\color{black}SPE-MF-TBD},  DBT, G-TBD algorithms in the whole surveillance area. }
	\label{fig: track_real_data}
\end{figure*}

The {\color{black}SPE-MF-TBD} method outputs target long tracks through a sliding-window-based batch processing structure using a fixed batch processing length $K$ for each window~\cite{WangTAES}.
The experimental results of the
existing widely used SFD, clustering, and EKF approaches are also given as a benchmark. Note that the traditional MF-TBD methods suffer from severe performance degradation due to neglecting the movement of the radar platform itself, as discussed in Section~\ref{sec:Results}. Thus, the traditional MF-TBD methods are not considered in this experiment.

 \begin{figure}
	\centering
\subfigure{\centering\includegraphics[width=3.4in]{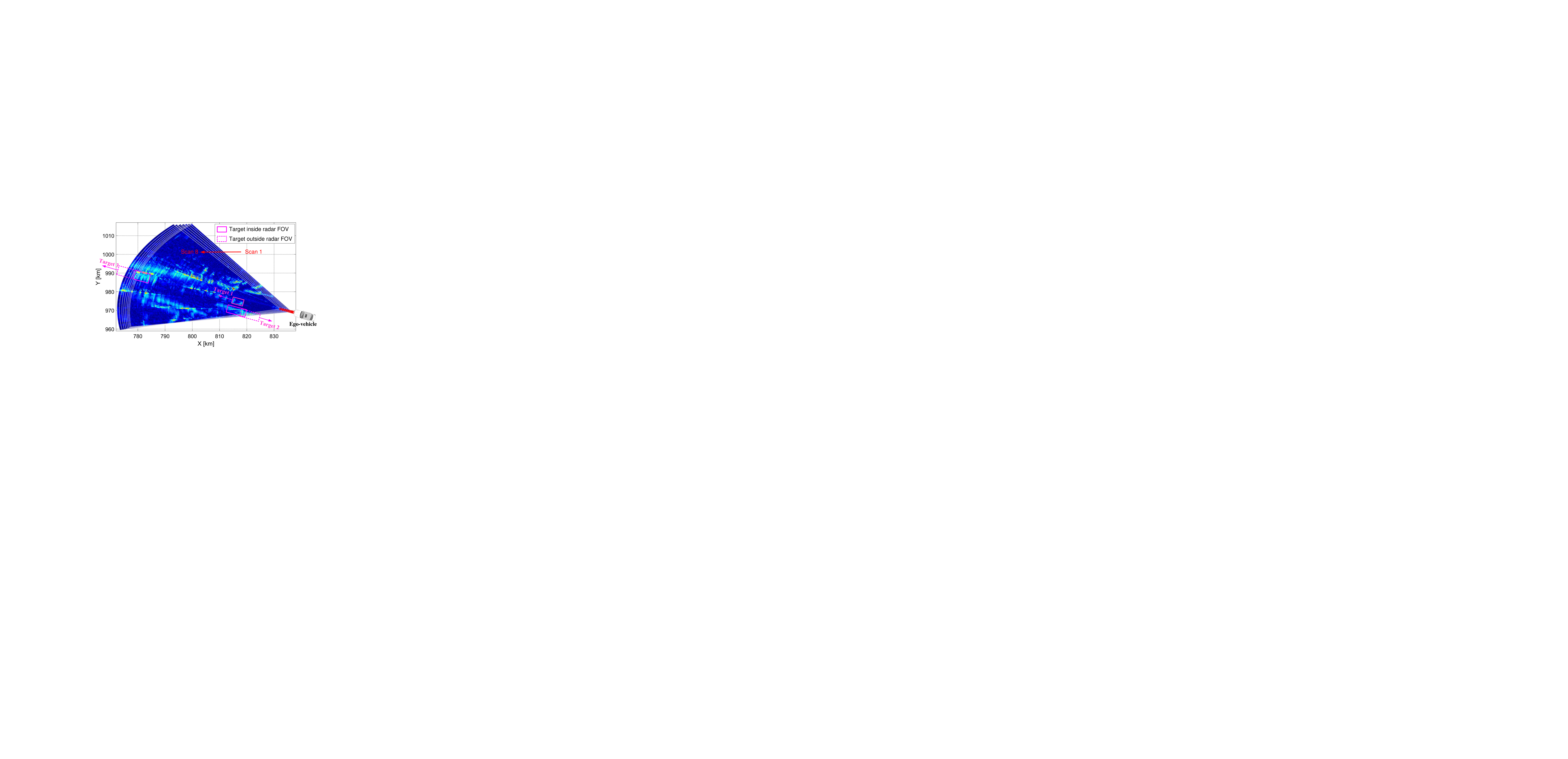}}
	\caption{{\color{black}Automotive radar echo measurements of an overlaid snapshot of eight scans. }
     }
	\label{fig: MEA_RADAR_FOV}
\end{figure}

The optimal search paths produced by the {\color{black}SPE-MF-TBD and G-TBD} algorithm can be declared after running the multi-frame integration and detection procedures for $K$--scan data of each processing window. A zoomed-in view for estimated tracks of targets $1$--$3$ in a batch processing windows with scans $1$--$6$ is shown in Fig.~\ref{fig: zone_track}, where raw echo measurements are plotted as a background to clearly show the relationship between the echo measurements and the estimated tracks. The black points in Fig.~\ref{fig: zone_track} are the result of SFD and clustering processes.
It is shown in Fig.~\ref{fig: zone_track} that the proposed {\color{black}SPE-MF-TBD} algorithm can accurately integrate target energy along the target motion path from echo measurements of a moving platform radar.
{\color{black} However, the existing G-TBD and SFD methods experience target information loss after threshold detection, making it challenging to track target $1$ with lower SNR, as shown in Fig.~\ref{fig: zone_track}.
} In practice, to avoid declaring a large number of false measurement points, SFD usually selects a higher detection threshold value, which inevitably leads to the loss of weak targets.  

All estimated tracks of {\color{black}SPE-MF-TBD and G-TBD} are declared via a clustering operation to merge different sliding window signals that appear to come from the same target in Fig.~\ref{fig: track_real_data}. Similarly, the DBT tracking results are obtained after processing the data association and filtering of all point measurements. As expected, the {\color{black}SPE-MF-TBD} algorithm outputs more complete and accurate target tracks. In addition, the number of false tracks produced by the {\color{black}SPE-MF-TBD} approach is less than that of DBT. The DBT method fails to track targets when the target's SNR is low. For example, only the target~$2$ with sufficiently strong energy is tracked in its entirety, the track of target 3 is fractured into multiple parts, and target 1  is almost completely lost in Fig.~\ref{fig: track_real_data}. {\color{black}We observe that G-TBD experiences model mismatch when tracking turning targets, making it challenging to maintain a long track for target 2, even under high SNR conditions.

\begin{figure}
	\centering
\subfigure{\centering\includegraphics[width=1.71in]{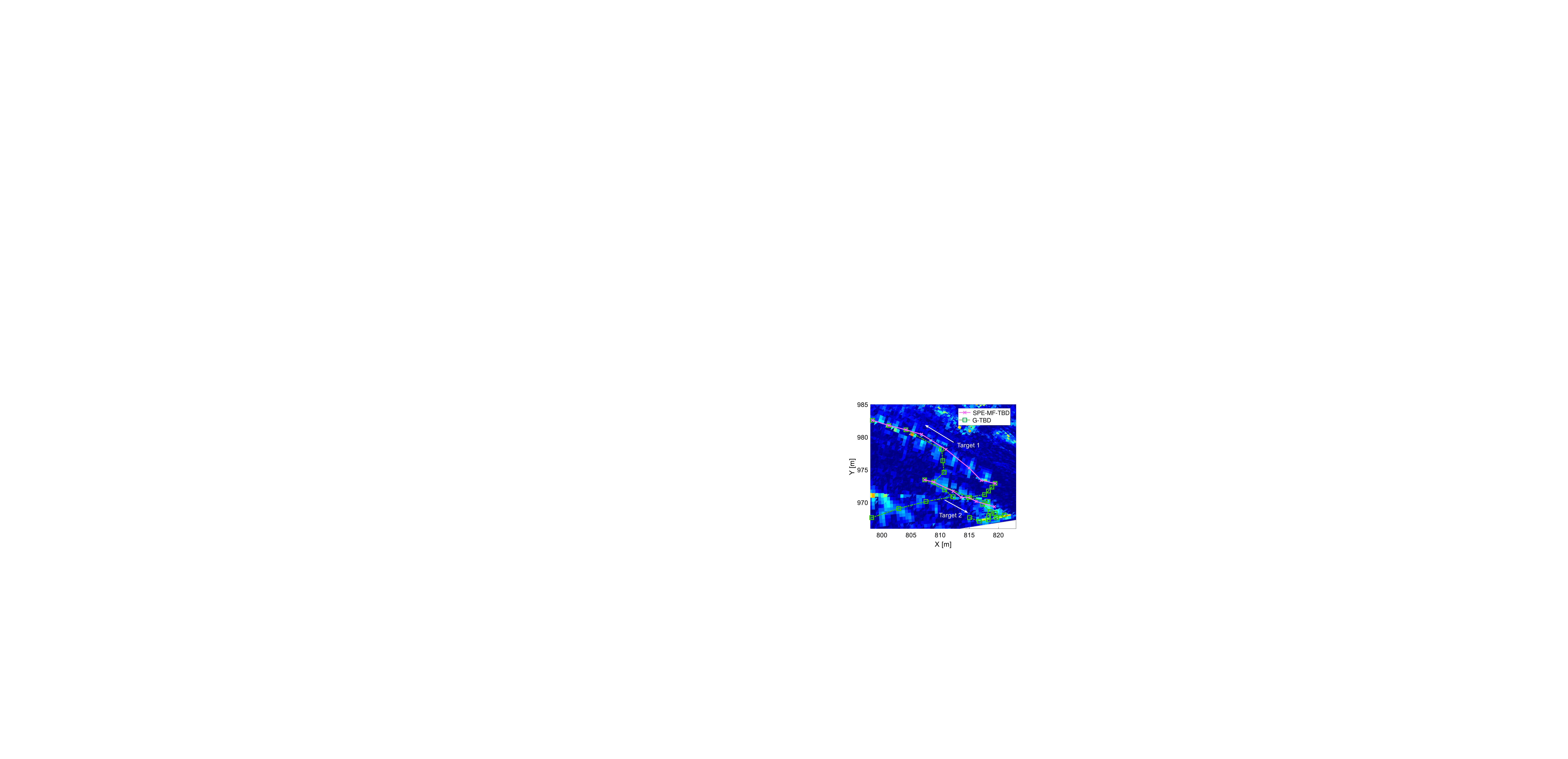}}
	\subfigure{\centering\includegraphics[width=1.73in]{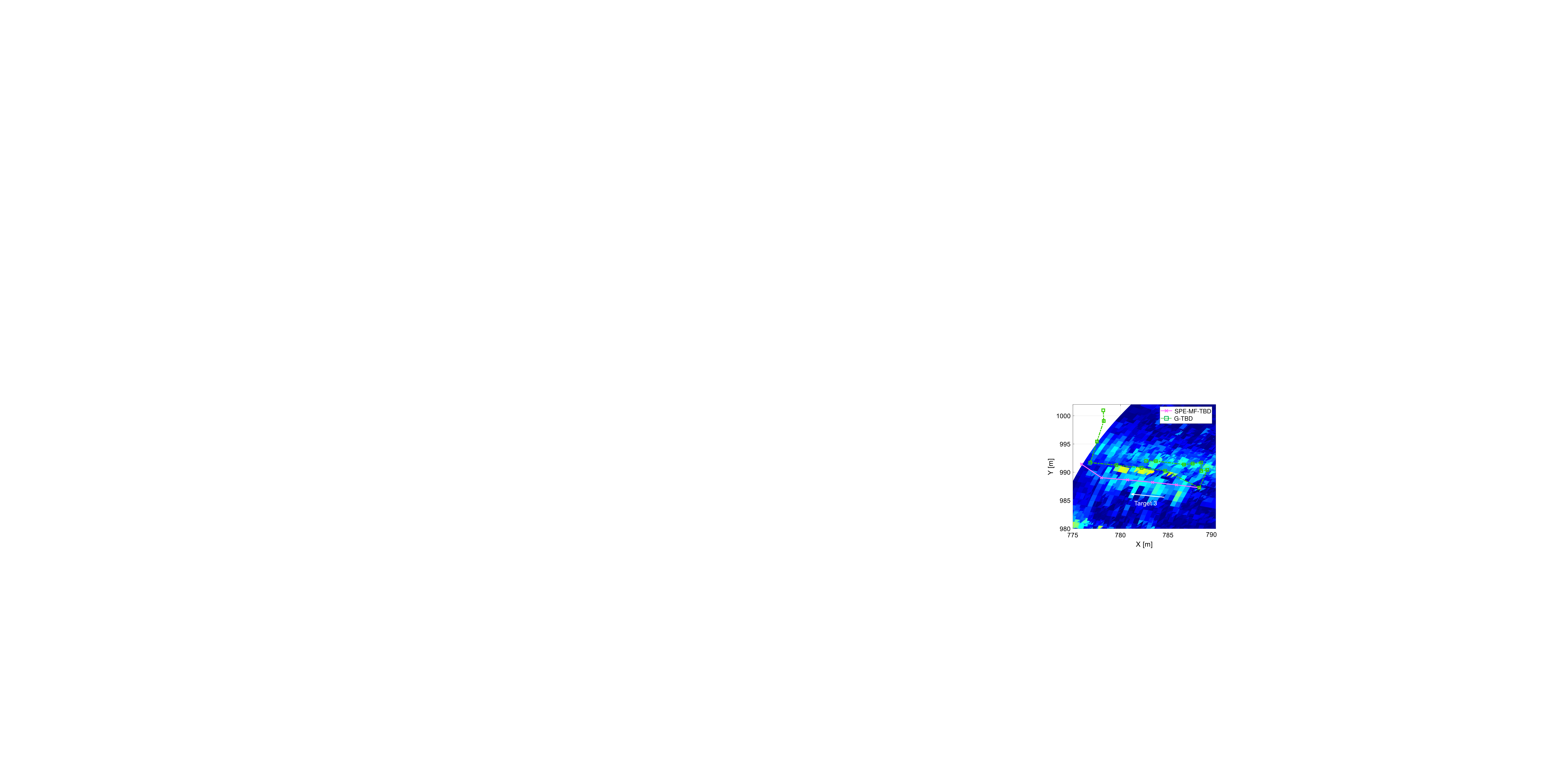}}
	\caption{Optimal search paths produced by SPE-MF-TBD and G-TBD methods with $K=8$. 
 }
	\label{fig: zone_track_FINAL}
\end{figure}

To further validate the efficiency of the proposed algorithm, a more complex experimental scenario is presented in Fig.~\ref{fig: MEA_RADAR_FOV}, where targets exit radar FOV during $K=8$ scans. Targets $2$ and $3$ can be observed moving near the edge of the radar FOV, eventually moving out of the radar FOV at $K=6$ and $K=7$. In contrast, the target $1$ remains within the radar FOV. The corresponding tracking results of SPE-MF-TBD and G-TBD are presented in Fig.~\ref{fig: zone_track_FINAL}. 
For targets $1$, $2$, and $3$, SPE-MF-TBD consistently exhibits superior robustness in accurately integrating energies along the target's paths. As expected, SPE-MF-TBD can accurately detect the length of the target track within radar FOV, owing to its adaptive detection architecture described in Section~\ref{se:detection}. However, G-TBD typically employs a constant detection threshold to extract targets, which can result in target loss, especially when the target moves beyond the radar FOV. Moreover, when the target is outside the FOV, G-TBD is forced to correlate with noise measurements, increasing the risk of model mismatch and simultaneously reducing tracking accuracy, as demonstrated by targets $2$ and $3$ in Fig.~\ref{fig: zone_track_FINAL}.
}

\section{Conclusion}\label{sec:Conclusion}
In this work, we have investigated MF-TBD procedures in a vehicle-mounted moving platform radar system. A novel 
MF-TBD detection architecture has been presented for automotive radar. It can adaptively adjust the detection threshold value based on the existence of a moving target within radar FOV. Then, an implementation of automotive radar MF-TBD, which accurately describes the nonlinear map relationships between different coordinates and measurement errors of sensors, has been derived for search path design. The self-positioning errors of the ego-vehicle, which are usually not considered in some previous target tracking approaches, have been well calibrated. Additionally, numerical simulation and experimental results with real data have shown that the proposed approaches guarantee a large detection and tracking gain with respect to the existing MF-TBD, G-TBD, and DBT processing in weak target detection scenarios. {\color{black} In future studies, we will further explore multi-frame integration approaches to address complex non-Gaussian localization errors and evaluate various localization methods. Additionally, we will investigate fast implementation techniques to enhance the practical application of the proposed algorithms in automotive radar systems.}

\begin{appendices}
\section{Conversion Error of the Yaw Angle} \label{Apen: cov}
 Recall that ${\color{black}\mathbf{x}}^\text{V}=\overline{{\color{black}\mathbf{x}}}^\text{V}+{\color{black}\mathbf{v}}^\text{V}$ in  (\ref{eq:CV_STATE_grid}), which is the grid state after compensating the mounting angle. Its mean $\bm{\mu}^\text{V}\in{\mathbb{R}^{4\times1}}$ and the conversion matrix $\bm{Q}^\text{V}\in{\mathbb{R}^{4\times4}}$ of the error ${\color{black}\mathbf{v}}^\text{V}$ are given in (\ref{eq:mean_cov1_1}) and (\ref{eq:mean_cov1_2}), respectively. The measured yaw angle of the ego-vehicle satisfies 
 $\alpha=\bar{\alpha}+{\omega}_{\alpha}$ with ${\omega}_{\alpha}\sim\mathcal{N}(\bm{0},\bm{Q}_{55}^\text{p})$ according to (\ref{eq:Vehicle_STATE}). For simplicity, the index  $^\text{V}$ of the superscript in ${\color{black}\mathbf{x}}^\text{V}$, $\bm{\mu}^\text{V}$ and $\bm{Q}^\text{V}$ has
been omitted. Then, after some algebraic manipulations, the specific expression of each element in the conversion matrix $\bm{Q}^\text{Y}\in{\mathbb{R}^{4\times4}}$ of (\ref{eq:carti_cov}) can be calculated as follows:
\begin{eqnarray}
{Q}_{11}^\text{Y}\hspace{-1.5mm}&=&\hspace{-1.5mm}\sin(2\alpha)\left(\lambda_{\text{C}}\hspace{-0.5mm}-\hspace{-0.5mm}\lambda_{\text{D}}\right)\big(xy-x\bm{\mu}_3-y\bm{\mu}_1+\bm{Q}_{13}\notag\\
&&\hspace{-1.5mm}+\bm{\mu}_1\bm{\mu}_3\big)-\sin\left(2\alpha\right)\lambda_{\text{B}}^2(x-\bm{\mu}_1)(y-\bm{\mu}_3)\notag\\
&&\hspace{-1.5mm}+(x^2\hspace{-0.5mm}+\hspace{-0.5mm}\bm{Q}_{11}\hspace{-0.5mm}+\hspace{-0.5mm}\bm{\mu}_1^2\hspace{-0.5mm}-\hspace{-0.5mm}2x\bm{\mu}_1)(\lambda_{\text{C}}\cos^2\alpha\hspace{-0.5mm}+\hspace{-0.5mm}\lambda_{\text{D}}\sin^2\alpha)\notag\\
&&\hspace{-1.5mm}+(y^2\hspace{-0.5mm}+\hspace{-0.5mm}\bm{Q}_{33}\hspace{-0.5mm}+\hspace{-0.5mm}\bm{\mu}_3^2\hspace{-0.5mm}-\hspace{-0.5mm}2y\bm{\mu}_3)(\lambda_{\text{C}}\sin^2\alpha\hspace{-0.5mm}+\hspace{-0.5mm}\lambda_{\text{D}}\cos^2\alpha)\notag\\
&&\hspace{-1.5mm}-\lambda_{\text{B}}^2(x-\bm{\mu}_1)^2\cos^2\alpha-\lambda_{\text{B}}^2(y-\bm{\mu}_3)^2\sin^2\alpha\\
{Q}_{22}^\text{Y}\hspace{-1.5mm}&=&\hspace{-1.5mm}\sin(2\alpha)(\lambda_{\text{C}}\hspace{-0.5mm}-\hspace{-0.5mm}\lambda_{\text{D}})\big(\dot{x}\dot{y}-\dot{x}\bm{\mu}_4-\dot{y}\bm{\mu}_2+\bm{Q}_{24}\notag\\
&&\hspace{-1.5mm}+\bm{\mu}_2\bm{\mu}_4\big)-\sin(2\alpha)\lambda_{\text{B}}^2(\dot{x}-\bm{\mu}_2)(\dot{y}-\bm{\mu}_4)\notag\\
&&\hspace{-1.5mm}+(\dot{x}^2+\bm{Q}_{22}+\bm{\mu}_2^2-2\dot{x}\bm{\mu}_2)(\lambda_{\text{C}}\cos^2\alpha+\lambda_{\text{D}}\sin^2\alpha)\notag\\
&&\hspace{-1.5mm}+(\dot{y}^2+\bm{Q}_{44}+\bm{\mu}_4^2-2\dot{y}\bm{\mu}_4)(\lambda_{\text{C}}\sin^2\alpha+\lambda_{\text{D}}\cos^2\alpha)\notag\\
&&\hspace{-1.5mm}-\lambda_{\text{B}}^2(\dot{x}-\bm{\mu}_2)^2\cos^2\alpha-\lambda_{\text{B}}^2(\dot{y}-\bm{\mu}_4)^2\sin^2\alpha
\end{eqnarray}
\begin{eqnarray}
{Q}_{33}^\text{Y}\hspace{-1.5mm}&=&\hspace{-1.5mm}\sin(2\alpha)\lambda_{\text{B}}^2(x-\bm{\mu}_1)(y-\bm{\mu}_3)-\sin(2\alpha)(\lambda_{\text{C}}\hspace{-0.5mm}-\hspace{-0.5mm}\lambda_{\text{D}})\notag\\
&&\hspace{-1.5mm}\times(xy-x\bm{\mu}_3-y\bm{\mu}_1+\bm{Q}_{13}+\bm{\mu}_1\bm{\mu}_3)\notag\\
&&\hspace{-1.5mm}+(x^2+\bm{Q}_{11}+\bm{\mu}_1^2-2x\bm{\mu}_1)(\lambda_{\text{C}}\sin^2\alpha+\lambda_{\text{D}}\cos^2\alpha)\notag\\
&&\hspace{-1.5mm}+(y^2+\bm{Q}_{33}+\bm{\mu}_3^2-2y\bm{\mu}_3)(\lambda_{\text{C}}\cos^2\alpha+\lambda_{\text{D}}\sin^2\alpha)\notag\\
&&\hspace{-1.5mm}-\lambda_{\text{B}}^2(x-\bm{\mu}_1)^2\sin^2\alpha-\lambda_{\text{B}}^2(y-\bm{\mu}_3)^2\cos^2\alpha\\
{Q}_{44}^\text{Y}\hspace{-1.5mm}&=&\hspace{-1.5mm}\sin(2\alpha)\lambda_{\text{B}}^2(\dot{x}-\bm{\mu}_2)(\dot{y}-\bm{\mu}_4)-\sin(2\alpha)(\lambda_{\text{C}}\hspace{-0.5mm}-\hspace{-0.5mm}\lambda_{\text{D}})\notag\\
&&\hspace{-1.5mm}\times(\dot{x}\dot{y}-\dot{x}\bm{\mu}_4-\dot{y}\bm{\mu}_2+\bm{Q}_{24}+\bm{\mu}_2\bm{\mu}_4)\notag\\
&&\hspace{-1.5mm}+(\dot{x}^2+\bm{Q}_{22}+\bm{\mu}_2^2-2\dot{x}\bm{\mu}_2)(\lambda_{\text{C}}\sin^2\alpha+\lambda_{\text{D}}\cos^2\alpha)\notag\\
&&\hspace{-1.5mm}+(\dot{y}^2+\bm{Q}_{44}+\bm{\mu}_4^2-2\dot{y}\bm{\mu}_4)(\lambda_{\text{C}}\cos^2\alpha+\lambda_{\text{D}}\sin^2\alpha)\notag\\
&&\hspace{-1.5mm}-\lambda_{\text{B}}^2(\dot{x}-\bm{\mu}_1)^2\sin^2\alpha-\lambda_{\text{B}}^2(\dot{y}-\bm{\mu}_3)^2\cos^2\alpha\\
{Q}_{12}^\text{Y}\hspace{-1.5mm}&=&\hspace{-1.5mm}(x\dot{x}+\bm{Q}_{12}+\bm{\mu}_1\bm{\mu}_2-\dot{x}\bm{\mu}_1-{x}\bm{\mu}_2)\big(\lambda_{\text{C}}\cos^2\alpha\notag\\
&&\hspace{-1.5mm}+\lambda_{\text{D}}\sin^2\alpha\big)+(\lambda_{\text{C}}\hspace{-0.5mm}-\hspace{-0.5mm}\lambda_{\text{D}})/2\sin(2\alpha)\big(\dot{x}y+\bm{Q}_{23}\notag\\
&&\hspace{-1.5mm}+\bm{\mu}_2\bm{\mu}_3-\dot{x}\bm{\mu}_3-y\bm{\mu}_2+x\dot{y}+\bm{Q}_{14}+\bm{\mu}_1\bm{\mu}_4-\dot{y}\bm{\mu}_1\notag\\
&&\hspace{-1.5mm}-x\bm{\mu}_4\big)+(y\dot{y}+\bm{Q}_{34}+\bm{\mu}_3\bm{\mu}_4-\dot{y}\bm{\mu}_3-{y}\bm{\mu}_4)\notag\\
&&\hspace{-1.5mm}\times(\lambda_{\text{C}}\sin^2\alpha+\lambda_{\text{D}}\cos^2\alpha)-\lambda_{\text{B}}^2\big[(x-\bm{\mu}_1)\cos\alpha+\notag\\
&&\hspace{-1.5mm}(y\hspace{-0.5mm}-\hspace{-0.5mm}\bm{\mu}_3)\sin\alpha\big] \big[(\dot{x}\hspace{-0.5mm}-\hspace{-0.5mm}\bm{\mu}_2)\cos\alpha\hspace{-0.5mm}+\hspace{-0.5mm}(\dot{y}-\bm{\mu}_4)\sin\alpha\big]
\\
{Q}_{13}^\text{Y}\hspace{-1.5mm}&=&\hspace{-1.5mm}(xy+\bm{Q}_{13}+\bm{\mu}_1\bm{\mu}_3-y\bm{\mu}_1-{x}\bm{\mu}_3)(\lambda_{\text{C}}-\lambda_{\text{D}})\cos(2\alpha)\notag\\
&&\hspace{-1.5mm}+(\lambda_{\text{C}}\hspace{-0.5mm}-\hspace{-0.5mm}\lambda_{\text{D}})/2\sin(2\alpha)\big(y^2+\bm{Q}_{33}+\bm{\mu}_3^2-2y\bm{\mu}_3\notag\\
&&\hspace{-1.5mm}-x^2-\bm{Q}_{11}-\bm{\mu}_1^2+2x\bm{\mu}_1\big)-\lambda_{\text{B}}^2\big[(x-\bm{\mu}_1)\cos\alpha+\notag\\
&&\hspace{-1.5mm}(y\hspace{-0.5mm}-\hspace{-0.5mm}\bm{\mu}_3)\sin\alpha\big] \big[(y\hspace{-0.5mm}-\hspace{-0.5mm}\bm{\mu}_3)\cos\alpha\hspace{-0.5mm}-\hspace{-0.5mm}(x-\bm{\mu}_1)\sin\alpha\big]\\
{Q}_{14}^\text{Y}\hspace{-1.5mm}&=&\hspace{-1.5mm}(x\dot{y}+\bm{Q}_{14}+\bm{\mu}_1\bm{\mu}_4-x\bm{\mu}_4-\dot{y}\bm{\mu}_1)\big(\lambda_{\text{C}}\cos^2\alpha\notag\\
&&\hspace{-1.5mm}+\lambda_{\text{D}}\sin^2\alpha\big)+(\lambda_{\text{C}}\hspace{-0.5mm}-\hspace{-0.5mm}\lambda_{\text{D}})/2\sin(2\alpha)\big(y\dot{y}+\bm{Q}_{34}\notag\\
&&\hspace{-1.5mm}+\bm{\mu}_3\bm{\mu}_4-y\bm{\mu}_4-\dot{y}\bm{\mu}_3-x\dot{x}-\bm{Q}_{12}-\bm{\mu}_1\bm{\mu}_2+x\bm{\mu}_2\notag\\
&&\hspace{-1.5mm}+\dot{x}\bm{\mu}_1\big)-(y\dot{x}+\bm{Q}_{23}+\bm{\mu}_2\bm{\mu}_3-\dot{x}\bm{\mu}_3-{y}\bm{\mu}_2)\notag\\
&&\hspace{-1.5mm}\times(\lambda_{\text{C}}\sin^2\alpha+\lambda_{\text{D}}\cos^2\alpha)-\lambda_{\text{B}}^2\big[(x-\bm{\mu}_1)\cos\alpha+\notag\\
&&\hspace{-1.5mm}(y\hspace{-0.5mm}-\hspace{-0.5mm}\bm{\mu}_3)\sin\alpha\big] \big[(\dot{y}\hspace{-0.5mm}-\hspace{-0.5mm}\bm{\mu}_4)\cos\alpha\hspace{-0.5mm}-\hspace{-0.5mm}(\dot{x}\hspace{-0.5mm}-\hspace{-0.5mm}\bm{\mu}_2)\sin\alpha\big]\\
{Q}_{23}^\text{Y}\hspace{-1.5mm}&=&\hspace{-1.5mm}(y\dot{x}+\bm{Q}_{23}+\bm{\mu}_2\bm{\mu}_3-y\bm{\mu}_2-\dot{x}\bm{\mu}_3)(\lambda_{\text{C}}\sin^2\alpha\notag\\
&&\hspace{-1.5mm}+\lambda_{\text{D}}\cos^2\alpha)+(\lambda_{\text{C}}\hspace{-0.5mm}-\hspace{-0.5mm}\lambda_{\text{D}})/2\sin(2\alpha)(y\dot{y}+\bm{Q}_{34}\notag\\
&&\hspace{-1.5mm}+\bm{\mu}_3\bm{\mu}_4-y\bm{\mu}_4-\dot{y}\bm{\mu}_3-x\dot{x}-\bm{Q}_{12}-\bm{\mu}_1\bm{\mu}_2+x\bm{\mu}_2\notag\\
&&\hspace{-1.5mm}+\dot{x}\bm{\mu}_1)-(x\dot{y}+\bm{Q}_{14}+\bm{\mu}_1\bm{\mu}_4-x\bm{\mu}_4-\dot{y}\bm{\mu}_1)\notag\\
&&\hspace{-1.5mm}\times(\lambda_{\text{C}}\sin^2\alpha+\lambda_{\text{D}}\cos^2\alpha)-\lambda_{\text{B}}^2\big[(\dot{x}-\bm{\mu}_2)\cos\alpha+\notag\\
&&\hspace{-1.5mm}(\dot{y}\hspace{-0.5mm}-\hspace{-0.5mm}\bm{\mu}_4)\sin\alpha\big]\big[({y}\hspace{-0.5mm}-\hspace{-0.5mm}\bm{\mu}_3)\cos\alpha\hspace{-0.5mm}-\hspace{-0.5mm}({x}\hspace{-0.5mm}-\hspace{-0.5mm}\bm{\mu}_1)\sin\alpha\big]\\
{Q}_{24}^\text{Y}\hspace{-1.5mm}&=&\hspace{-1.5mm}(\dot{x}\dot{y}-\dot{x}\bm{\mu}_4-\dot{y}\bm{\mu}_2+\bm{Q}_{24}+\bm{\mu}_2\bm{\mu}_4)(\lambda_{\text{C}}-\lambda_{\text{D}})\cos(2\alpha)\notag\\
&&\hspace{-1.5mm}+(\lambda_{\text{C}}\hspace{-0.5mm}-\hspace{-0.5mm}\lambda_{\text{D}})/2\sin(2\alpha)\big(\dot{y}^2+\bm{Q}_{44}+\bm{\mu}_4^2-2\dot{y}\bm{\mu}_4\notag\\
&&\hspace{-1.5mm}-\dot{x}^2-\bm{Q}_{22}-\bm{\mu}_2^2+2\dot{x}\bm{\mu}_2\big)-\lambda_{\text{B}}^2\big[(\dot{x}-\bm{\mu}_2)\cos\alpha+\notag\\
&&\hspace{-1.5mm}(\dot{y}\hspace{-0.5mm}-\hspace{-0.5mm}\bm{\mu}_4)\sin\alpha\big]\big[(\dot{y}\hspace{-0.5mm}-\hspace{-0.5mm}\bm{\mu}_4)\cos\alpha\hspace{-0.5mm}-\hspace{-0.5mm}(\dot{x}\hspace{-0.5mm}-\hspace{-0.5mm}\bm{\mu}_2)\sin\alpha\big]\\
{Q}_{34}^\text{Y}\hspace{-1.5mm}&=&\hspace{-1.5mm}(x\dot{x}+\bm{Q}_{12}+\bm{\mu}_1\bm{\mu}_2-\dot{x}\bm{\mu}_1-{x}\bm{\mu}_2)\big(\lambda_{\text{C}}\sin^2\alpha\notag\\
&&\hspace{-1.5mm}+\lambda_{\text{D}}\cos^2\alpha\big)+(y\dot{y}+\bm{Q}_{34}+\bm{\mu}_3\bm{\mu}_4-\dot{y}\bm{\mu}_3-{y}\bm{\mu}_4)\notag\\
&&\hspace{-1.5mm}\times(\lambda_{\text{C}}\cos^2\alpha+\lambda_{\text{D}}\sin^2\alpha)-(\lambda_{\text{C}}\hspace{-0.5mm}-\hspace{-0.5mm}\lambda_{\text{D}})/2\sin(2\alpha)\notag\\
&&\hspace{-1.5mm}\times\big(x\dot{y}+\bm{Q}_{14}+\bm{\mu}_1\bm{\mu}_4-x\bm{\mu}_4-\dot{y}\bm{\mu}_1+y\dot{x}+\bm{Q}_{23}\notag\\
&&\hspace{-1.5mm}+\bm{\mu}_2\bm{\mu}_3-y\bm{\mu}_2-\dot{x}\bm{\mu}_3\big)-\lambda_{\text{B}}^2\big[(y-\bm{\mu}_3)\cos\alpha-\notag\\
&&\hspace{-1.5mm}(x\hspace{-0.5mm}-\hspace{-0.5mm}\bm{\mu}_1)\sin\alpha\big]\big[(\dot{y}\hspace{-0.5mm}-\hspace{-0.5mm}\bm{\mu}_4)\cos\alpha\hspace{-0.5mm}-\hspace{-0.5mm}(\dot{x}\hspace{-0.5mm}-\hspace{-0.5mm}\bm{\mu}_2)\sin\alpha\big]
\end{eqnarray}
with 
\begin{align}
&{Q}_{21}^\text{Y}={Q}_{12}^\text{Y}, &{Q}_{31}^\text{Y}={Q}_{13}^\text{Y}, \hspace{8mm} {Q}_{41}^\text{Y}={Q}_{14}^\text{Y},\\
&{Q}_{32}^\text{Y}={Q}_{23}^\text{Y}, &{Q}_{42}^\text{Y}={Q}_{24}^\text{Y},  \hspace{8mm}{Q}_{43}^\text{Y}={Q}_{34}^\text{Y},
\end{align}
where $\lambda_{\text{C}}=(1+\exp^{-2{Q}^\text{p}_{55}})/2$ and $\lambda_{\text{D}}=(1-\exp^{-2{Q}^\text{p}_{55}})/2$.
\end{appendices}

\bibliographystyle{ieeetr}

\bibliography{references}

\end{document}